\documentclass[longauth]{aa}
\usepackage{txfonts}
\usepackage{amsmath}
\usepackage{xcolor}
\usepackage{eht}
\usepackage[colorlinks, allcolors=blue, breaklinks=true]{hyperref}

\addto\extrasenglish{

}

\makeatletter
\renewcommand*\maketitle{
  \thispagestyle{firstpage}
\begingroup
    \if@wideboxfn
    \setlength\bibindent{1.4\parindent}
    \else
    \setlength\bibindent{\parindent}
    \fi
    \renewcommand*\thefootnote{\@fnsymbol\c@footnote}
    \renewcommand\@makefntext[1]{
    \ifaa@longfn\hsize\textwidth\fi
    \noindent
    \hb@xt@\bibindent{\hss\@makefnmark\enspace}##1}
  \ifaa@twocolumn
  \begin{aa@strip}
    \aa@maketitle
    \@thanks
  \end{aa@strip}
  \else
    \begingroup
      \let\thanks\footnote
      \aa@maketitle
    \endgroup
  \fi
\endgroup
  \setcounter{footnote}{0}
}

\renewcommand*\aa@pageof{, page \thepage{} of \pageref*{LastPage}}
\makeatother

\usepackage{graphicx}
\usepackage{comment}
\usepackage[encapsulated]{CJK}
\usepackage{ucs}
\usepackage{txfonts}
\newcommand{\orcid}[1]{\protect\href{https://orcid.org/#1}{\protect\includegraphics[width=8pt]{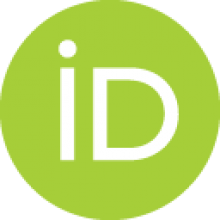}}}

\usepackage[utf8]{inputenc}
\newcommand{\cntext}[1]{\begin{CJK}{UTF8}{gbsn}#1\end{CJK}}

\begin{document}

\title{Locating the missing large-scale emission in the jet of M87* with short EHT baselines}

\author{
Boris Georgiev\orcid{0000-0002-3586-6424}\inst{\ref{inst14}}\and
Paul Tiede\orcid{0000-0003-3826-5648}\inst{\ref{inst10},\ref{inst3}}\and
Sebastiano D. von Fellenberg\orcid{0000-0002-9156-2249}\inst{\ref{inst131},\ref{inst6}}\and
Michael Janssen\orcid{0000-0001-8685-6544}\inst{\ref{inst29},\ref{inst6}}\and
Iniyan Natarajan\orcid{0000-0001-8242-4373}\inst{\ref{inst10},\ref{inst3}}\and
Lindy Blackburn\orcid{0000-0002-9030-642X}\inst{\ref{inst10},\ref{inst3}}\and
Jongho Park\orcid{0000-0001-6558-9053}\inst{\ref{inst130},\ref{inst11}}\and
Erandi Chavez\orcid{0000-0003-4143-9717}\inst{\ref{inst10}}\and
Andrew T. West\orcid{0000-0002-5471-4709}\inst{\ref{inst14}}\and
Kotaro Moriyama\orcid{0000-0003-1364-3761}\inst{\ref{inst47},\ref{inst79}}\and
Jun Yi Koay\orcid{0000-0002-7029-6658}\inst{\ref{inst101},\ref{inst11}}\and
Hendrik Müller\orcid{0000-0002-9250-0197}\inst{\ref{inst6}}\and
Dhanya G. Nair\orcid{0000-0001-5357-7805}\inst{\ref{inst16},\ref{inst6}}\and
Avery E. Broderick\orcid{0000-0002-3351-760X}\inst{\ref{inst26},\ref{inst27},\ref{inst28}}\and
Maciek Wielgus\orcid{0000-0002-8635-4242}\inst{\ref{inst5}}
\\The Event Horizon Telescope Collaboration:\\
Kazunori Akiyama\orcid{0000-0002-9475-4254}\inst{\ref{inst1},\ref{inst2},\ref{inst3}}\and
Ezequiel Albentosa-Ruíz\orcid{0000-0002-7816-6401}\inst{\ref{inst4}}\and
Antxon Alberdi\orcid{0000-0002-9371-1033}\inst{\ref{inst5}}\and
Walter Alef\inst{\ref{inst6}}\and
Juan Carlos Algaba\orcid{0000-0001-6993-1696}\inst{\ref{inst7}}\and
Richard Anantua\orcid{0000-0003-3457-7660}\inst{\ref{inst8},\ref{inst9},\ref{inst3},\ref{inst10}}\and
Keiichi Asada\orcid{0000-0001-6988-8763}\inst{\ref{inst11}}\and
Rebecca Azulay\orcid{0000-0002-2200-5393}\inst{\ref{inst4},\ref{inst12},\ref{inst6}}\and
Uwe Bach\orcid{0000-0002-7722-8412}\inst{\ref{inst6}}\and
Anne-Kathrin Baczko\orcid{0000-0003-3090-3975}\inst{\ref{inst13},\ref{inst6}}\and
David Ball\inst{\ref{inst14}}\and
Mislav Baloković\orcid{0000-0003-0476-6647}\inst{\ref{inst15}}\and
Bidisha Bandyopadhyay\orcid{0000-0002-2138-8564}\inst{\ref{inst16}}\and
John Barrett\orcid{0000-0002-9290-0764}\inst{\ref{inst1}}\and
Michi Bauböck\orcid{0000-0002-5518-2812}\inst{\ref{inst17}}\and
Bradford A. Benson\orcid{0000-0002-5108-6823}\inst{\ref{inst18},\ref{inst19}}\and
Dan Bintley\inst{\ref{inst20},\ref{inst21}}\and
Raymond Blundell\orcid{0000-0002-5929-5857}\inst{\ref{inst10}}\and
Katherine L. Bouman\orcid{0000-0003-0077-4367}\inst{\ref{inst22}}\and
Geoffrey C. Bower\orcid{0000-0003-4056-9982}\inst{\ref{inst20},\ref{inst21},\ref{inst23},\ref{inst24}}\and
Michael Bremer\inst{\ref{inst25}}\and
Roger Brissenden\orcid{0000-0002-2556-0894}\inst{\ref{inst10}}\and
Silke Britzen\orcid{0000-0001-9240-6734}\inst{\ref{inst6}}\and
Dominique Broguiere\orcid{0000-0001-9151-6683}\inst{\ref{inst25}}\and
Thomas Bronzwaer\orcid{0000-0003-1151-3971}\inst{\ref{inst29}}\and
Sandra Bustamante\orcid{0000-0001-6169-1894}\inst{\ref{inst30}}\and
Douglas F. Carlos\orcid{0000-0002-1340-7702}\inst{\ref{inst31}}\and
John E. Carlstrom\orcid{0000-0002-2044-7665}\inst{\ref{inst32},\ref{inst19},\ref{inst33},\ref{inst34}}\and
Andrew Chael\orcid{0000-0003-2966-6220}\inst{\ref{inst35}}\and
Chi-kwan Chan\orcid{0000-0001-6337-6126}\inst{\ref{inst14},\ref{inst36},\ref{inst37}}\and
Dominic O. Chang\orcid{0000-0001-9939-5257}\inst{\ref{inst10},\ref{inst3}}\and
Koushik Chatterjee\orcid{0000-0002-2825-3590}\inst{\ref{inst38},\ref{inst3},\ref{inst10}}\and
Shami Chatterjee\orcid{0000-0002-2878-1502}\inst{\ref{inst39}}\and
Ming-Tang Chen\orcid{0000-0001-6573-3318}\inst{\ref{inst23}}\and
Yongjun Chen (\cntext{陈永军})\orcid{0000-0001-5650-6770}\inst{\ref{inst40},\ref{inst41}}\and
Xiaopeng Cheng\orcid{0000-0003-4407-9868}\inst{\ref{inst42}}\and
Paul Chichura\orcid{0000-0002-5397-9035}\inst{\ref{inst33},\ref{inst32}}\and
Ilje Cho\orcid{0000-0001-6083-7521}\inst{\ref{inst42},\ref{inst43},\ref{inst5}}\and
Pierre Christian\orcid{0000-0001-6820-9941}\inst{\ref{inst44}}\and
Nicholas S. Conroy\orcid{0000-0003-2886-2377}\inst{\ref{inst45},\ref{inst10}}\and
John E. Conway\orcid{0000-0003-2448-9181}\inst{\ref{inst13}}\and
Thomas M. Crawford\orcid{0000-0001-9000-5013}\inst{\ref{inst19},\ref{inst32}}\and
Geoffrey B. Crew\orcid{0000-0002-2079-3189}\inst{\ref{inst1}}\and
Alejandro Cruz-Osorio\orcid{0000-0002-3945-6342}\inst{\ref{inst46},\ref{inst47}}\and
Yuzhu Cui (\cntext{崔玉竹})\orcid{0000-0001-6311-4345}\inst{\ref{inst48}}\and
Brandon Curd\orcid{0000-0002-8650-0879}\inst{\ref{inst8},\ref{inst3},\ref{inst10}}\and
Rohan Dahale\orcid{0000-0001-6982-9034}\inst{\ref{inst5}}\and
Jordy Davelaar\orcid{0000-0002-2685-2434}\inst{\ref{inst49},\ref{inst50}}\and
Mariafelicia De Laurentis\orcid{0000-0002-9945-682X}\inst{\ref{inst51},\ref{inst52}}\and
Roger Deane\orcid{0000-0003-1027-5043}\inst{\ref{inst53},\ref{inst54},\ref{inst55}}\and
Gregory Desvignes\orcid{0000-0003-3922-4055}\inst{\ref{inst6},\ref{inst56}}\and
Jason Dexter\orcid{0000-0003-3903-0373}\inst{\ref{inst57}}\and
Vedant Dhruv\orcid{0000-0001-6765-877X}\inst{\ref{inst17}}\and
Indu K. Dihingia\orcid{0000-0002-4064-0446}\inst{\ref{inst58}}\and
Sheperd S. Doeleman\orcid{0000-0002-9031-0904}\inst{\ref{inst10},\ref{inst3}}\and
Sergio A. Dzib\orcid{0000-0001-6010-6200}\inst{\ref{inst6}}\and
Ralph P. Eatough\orcid{0000-0001-6196-4135}\inst{\ref{inst59},\ref{inst6}}\and
Razieh Emami\orcid{0000-0002-2791-5011}\inst{\ref{inst10}}\and
Heino Falcke\orcid{0000-0002-2526-6724}\inst{\ref{inst29}}\and
Joseph Farah\orcid{0000-0003-4914-5625}\inst{\ref{inst60},\ref{inst61}}\and
Vincent L. Fish\orcid{0000-0002-7128-9345}\inst{\ref{inst1}}\and
Edward Fomalont\orcid{0000-0002-9036-2747}\inst{\ref{inst62}}\and
H. Alyson Ford\orcid{0000-0002-9797-0972}\inst{\ref{inst14}}\and
Marianna Foschi\orcid{0000-0001-8147-4993}\inst{\ref{inst5}}\and
Raquel Fraga-Encinas\orcid{0000-0002-5222-1361}\inst{\ref{inst29}}\and
William T. Freeman\inst{\ref{inst63},\ref{inst64}}\and
Per Friberg\orcid{0000-0002-8010-8454}\inst{\ref{inst20},\ref{inst21}}\and
Christian M. Fromm\orcid{0000-0002-1827-1656}\inst{\ref{inst65},\ref{inst47},\ref{inst6}}\and
Antonio Fuentes\orcid{0000-0002-8773-4933}\inst{\ref{inst5}}\and
Peter Galison\orcid{0000-0002-6429-3872}\inst{\ref{inst3},\ref{inst66},\ref{inst67}}\and
Charles F. Gammie\orcid{0000-0001-7451-8935}\inst{\ref{inst17},\ref{inst45},\ref{inst68}}\and
Roberto García\orcid{0000-0002-6584-7443}\inst{\ref{inst25}}\and
Olivier Gentaz\orcid{0000-0002-0115-4605}\inst{\ref{inst25}}\and
Ciriaco Goddi\orcid{0000-0002-2542-7743}\inst{\ref{inst31},\ref{inst69},\ref{inst70},\ref{inst71}}\and
Roman Gold\orcid{0000-0003-2492-1966}\inst{\ref{inst72},\ref{inst73},\ref{inst74}}\and
Arturo I. Gómez-Ruiz\orcid{0000-0001-9395-1670}\inst{\ref{inst75},\ref{inst76}}\and
José L. Gómez\orcid{0000-0003-4190-7613}\inst{\ref{inst5}}\and
Minfeng Gu (\cntext{顾敏峰})\orcid{0000-0002-4455-6946}\inst{\ref{inst40},\ref{inst77}}\and
Mark Gurwell\orcid{0000-0003-0685-3621}\inst{\ref{inst10}}\and
Kazuhiro Hada\orcid{0000-0001-6906-772X}\inst{\ref{inst78},\ref{inst79}}\and
Daryl Haggard\orcid{0000-0001-6803-2138}\inst{\ref{inst80},\ref{inst81}}\and
Ronald Hesper\orcid{0000-0003-1918-6098}\inst{\ref{inst82}}\and
Dirk Heumann\orcid{0000-0002-7671-0047}\inst{\ref{inst14}}\and
Luis C. Ho (\cntext{何子山})\orcid{0000-0001-6947-5846}\inst{\ref{inst83},\ref{inst84}}\and
Paul Ho\orcid{0000-0002-3412-4306}\inst{\ref{inst11},\ref{inst21},\ref{inst20}}\and
Mareki Honma\orcid{0000-0003-4058-9000}\inst{\ref{inst79},\ref{inst85},\ref{inst86}}\and
Chih-Wei L. Huang\orcid{0000-0001-5641-3953}\inst{\ref{inst11}}\and
Lei Huang (\cntext{黄磊})\orcid{0000-0002-1923-227X}\inst{\ref{inst40},\ref{inst77}}\and
David H. Hughes\inst{\ref{inst75}}\and
Shiro Ikeda\orcid{0000-0002-2462-1448}\inst{\ref{inst2},\ref{inst87},\ref{inst88},\ref{inst89}}\and
C. M. Violette Impellizzeri\orcid{0000-0002-3443-2472}\inst{\ref{inst90},\ref{inst62}}\and
Makoto Inoue\orcid{0000-0001-5037-3989}\inst{\ref{inst11}}\and
Sara Issaoun\orcid{0000-0002-5297-921X}\inst{\ref{inst10},\ref{inst50}}\and
David J. James\orcid{0000-0001-5160-4486}\inst{\ref{inst91},\ref{inst92}}\and
Buell T. Jannuzi\orcid{0000-0002-1578-6582}\inst{\ref{inst14}}\and
Britton Jeter\orcid{0000-0003-2847-1712}\inst{\ref{inst11}}\and
Wu Jiang (\cntext{江悟})\orcid{0000-0001-7369-3539}\inst{\ref{inst40}}\and
Alejandra Jiménez-Rosales\orcid{0000-0002-2662-3754}\inst{\ref{inst29}}\and
Michael D. Johnson\orcid{0000-0002-4120-3029}\inst{\ref{inst10},\ref{inst3}}\and
Svetlana Jorstad\orcid{0000-0001-6158-1708}\inst{\ref{inst93}}\and
Adam C. Jones\inst{\ref{inst19}}\and
Abhishek V. Joshi\orcid{0000-0002-2514-5965}\inst{\ref{inst17}}\and
Taehyun Jung\orcid{0000-0001-7003-8643}\inst{\ref{inst42},\ref{inst94}}\and
Ramesh Karuppusamy\orcid{0000-0002-5307-2919}\inst{\ref{inst6}}\and
Tomohisa Kawashima\orcid{0000-0001-8527-0496}\inst{\ref{inst95}}\and
Garrett K. Keating\orcid{0000-0002-3490-146X}\inst{\ref{inst10}}\and
Mark Kettenis\orcid{0000-0002-6156-5617}\inst{\ref{inst96}}\and
Dong-Jin Kim\orcid{0000-0002-7038-2118}\inst{\ref{inst97}}\and
Jae-Young Kim\orcid{0000-0001-8229-7183}\inst{\ref{inst98}}\and
Jongsoo Kim\orcid{0000-0002-1229-0426}\inst{\ref{inst42}}\and
Junhan Kim\orcid{0000-0002-4274-9373}\inst{\ref{inst99}}\and
Motoki Kino\orcid{0000-0002-2709-7338}\inst{\ref{inst2},\ref{inst100}}\and
Prashant Kocherlakota\orcid{0000-0001-7386-7439}\inst{\ref{inst3},\ref{inst10}}\and
Yutaro Kofuji\inst{\ref{inst79},\ref{inst86}}\and
Patrick M. Koch\orcid{0000-0003-2777-5861}\inst{\ref{inst11}}\and
Shoko Koyama\orcid{0000-0002-3723-3372}\inst{\ref{inst101},\ref{inst11}}\and
Carsten Kramer\orcid{0000-0002-4908-4925}\inst{\ref{inst25}}\and
Joana A. Kramer\orcid{0009-0003-3011-0454}\inst{\ref{inst6}}\and
Michael Kramer\orcid{0000-0002-4175-2271}\inst{\ref{inst6}}\and
Thomas P. Krichbaum\orcid{0000-0002-4892-9586}\inst{\ref{inst6}}\and
Cheng-Yu Kuo\orcid{0000-0001-6211-5581}\inst{\ref{inst102},\ref{inst11}}\and
Noemi La Bella\orcid{0000-0002-8116-9427}\inst{\ref{inst29}}\and
Deokhyeong Lee\orcid{0009-0003-2122-9437}\inst{\ref{inst103}}\and
Sang-Sung Lee\orcid{0000-0002-6269-594X}\inst{\ref{inst42}}\and
Aviad Levis\orcid{0000-0001-7307-632X}\inst{\ref{inst22}}\and
Shaoling Li\inst{\ref{inst20}}\and
Zhiyuan Li (\cntext{李志远})\orcid{0000-0003-0355-6437}\inst{\ref{inst104},\ref{inst105}}\and
Rocco Lico\orcid{0000-0001-7361-2460}\inst{\ref{inst106},\ref{inst5}}\and
Greg Lindahl\orcid{0000-0002-6100-4772}\inst{\ref{inst107}}\and
Michael Lindqvist\orcid{0000-0002-3669-0715}\inst{\ref{inst13}}\and
Mikhail Lisakov\orcid{0000-0001-6088-3819}\inst{\ref{inst108}}\and
Jun Liu (\cntext{刘俊})\orcid{0000-0002-7615-7499}\inst{\ref{inst6}}\and
Kuo Liu\orcid{0000-0002-2953-7376}\inst{\ref{inst40},\ref{inst41}}\and
Elisabetta Liuzzo\orcid{0000-0003-0995-5201}\inst{\ref{inst109}}\and
Wen-Ping Lo\orcid{0000-0003-1869-2503}\inst{\ref{inst11},\ref{inst110}}\and
Andrei P. Lobanov\orcid{0000-0003-1622-1484}\inst{\ref{inst6}}\and
Laurent Loinard\orcid{0000-0002-5635-3345}\inst{\ref{inst111},\ref{inst3},\ref{inst112}}\and
Colin J. Lonsdale\orcid{0000-0003-4062-4654}\inst{\ref{inst1}}\and
Amy E. Lowitz\orcid{0000-0002-4747-4276}\inst{\ref{inst14}}\and
Ru-Sen Lu (\cntext{路如森})\orcid{0000-0002-7692-7967}\inst{\ref{inst40},\ref{inst41},\ref{inst6}}\and
Nicholas R. MacDonald\orcid{0000-0002-6684-8691}\inst{\ref{inst6}}\and
Jirong Mao (\cntext{毛基荣})\orcid{0000-0002-7077-7195}\inst{\ref{inst113},\ref{inst114},\ref{inst115}}\and
Nicola Marchili\orcid{0000-0002-5523-7588}\inst{\ref{inst109},\ref{inst6}}\and
Sera Markoff\orcid{0000-0001-9564-0876}\inst{\ref{inst116},\ref{inst117}}\and
Daniel P. Marrone\orcid{0000-0002-2367-1080}\inst{\ref{inst14}}\and
Alan P. Marscher\orcid{0000-0001-7396-3332}\inst{\ref{inst93}}\and
Iván Martí-Vidal\orcid{0000-0003-3708-9611}\inst{\ref{inst4},\ref{inst12}}\and
Satoki Matsushita\orcid{0000-0002-2127-7880}\inst{\ref{inst11}}\and
Lynn D. Matthews\orcid{0000-0002-3728-8082}\inst{\ref{inst1}}\and
Lia Medeiros\orcid{0000-0003-2342-6728}\inst{\ref{inst118}}\and
Karl M. Menten\orcid{0000-0001-6459-0669}\inst{\ref{inst6},\ref{inst119}}\and
Izumi Mizuno\orcid{0000-0002-7210-6264}\inst{\ref{inst20},\ref{inst21}}\and
Yosuke Mizuno\orcid{0000-0002-8131-6730}\inst{\ref{inst58},\ref{inst120},\ref{inst47}}\and
Joshua Montgomery\orcid{0000-0003-0345-8386}\inst{\ref{inst81},\ref{inst19}}\and
James M. Moran\orcid{0000-0002-3882-4414}\inst{\ref{inst10},\ref{inst3}}\and
Monika Moscibrodzka\orcid{0000-0002-4661-6332}\inst{\ref{inst29}}\and
Wanga Mulaudzi\orcid{0000-0003-4514-625X}\inst{\ref{inst116}}\and
Cornelia Müller\orcid{0000-0002-2739-2994}\inst{\ref{inst6},\ref{inst29}}\and
Alejandro Mus\orcid{0000-0003-0329-6874}\inst{\ref{inst69},\ref{inst106},\ref{inst121},\ref{inst122}}\and
Gibwa Musoke\orcid{0000-0003-1984-189X}\inst{\ref{inst116},\ref{inst29}}\and
Ioannis Myserlis\orcid{0000-0003-3025-9497}\inst{\ref{inst123}}\and
Hiroshi Nagai\orcid{0000-0003-0292-3645}\inst{\ref{inst2},\ref{inst85}}\and
Neil M. Nagar\orcid{0000-0001-6920-662X}\inst{\ref{inst16}}\and
Masanori Nakamura\orcid{0000-0001-6081-2420}\inst{\ref{inst124},\ref{inst11}}\and
Gopal Narayanan\orcid{0000-0002-4723-6569}\inst{\ref{inst30}}\and
Antonios Nathanail\orcid{0000-0002-1655-9912}\inst{\ref{inst125},\ref{inst47}}\and
Santiago Navarro Fuentes\inst{\ref{inst123}}\and
Joey Neilsen\orcid{0000-0002-8247-786X}\inst{\ref{inst126}}\and
Chunchong Ni\orcid{0000-0003-1361-5699}\inst{\ref{inst27},\ref{inst28},\ref{inst26}}\and
Michael A. Nowak\orcid{0000-0001-6923-1315}\inst{\ref{inst127}}\and
Junghwan Oh\orcid{0000-0002-4991-9638}\inst{\ref{inst96}}\and
Hiroki Okino\orcid{0000-0003-3779-2016}\inst{\ref{inst79},\ref{inst86}}\and
Héctor Raúl Olivares Sánchez\orcid{0000-0001-6833-7580}\inst{\ref{inst128}}\and
Tomoaki Oyama\orcid{0000-0003-4046-2923}\inst{\ref{inst79}}\and
Feryal Özel\orcid{0000-0003-4413-1523}\inst{\ref{inst129}}\and
Daniel C. M. Palumbo\orcid{0000-0002-7179-3816}\inst{\ref{inst3},\ref{inst10}}\and
Georgios Filippos Paraschos\orcid{0000-0001-6757-3098}\inst{\ref{inst6}}\and
Harriet Parsons\orcid{0000-0002-6327-3423}\inst{\ref{inst20},\ref{inst21}}\and
Nimesh Patel\orcid{0000-0002-6021-9421}\inst{\ref{inst10}}\and
Ue-Li Pen\orcid{0000-0003-2155-9578}\inst{\ref{inst11},\ref{inst26},\ref{inst131},\ref{inst132},\ref{inst133}}\and
Dominic W. Pesce\orcid{0000-0002-5278-9221}\inst{\ref{inst10},\ref{inst3}}\and
Vincent Piétu\inst{\ref{inst25}}\and
Alexander Plavin\orcid{0000-0003-2914-8554}\inst{\ref{inst3},\ref{inst10},\ref{inst6}}\and
Aleksandar PopStefanija\inst{\ref{inst30}}\and
Oliver Porth\orcid{0000-0002-4584-2557}\inst{\ref{inst116},\ref{inst47}}\and
Ben Prather\orcid{0000-0002-0393-7734}\inst{\ref{inst17}}\and
Giacomo Principe\orcid{0000-0003-0406-7387}\inst{\ref{inst134},\ref{inst135},\ref{inst106}}\and
Dimitrios Psaltis\orcid{0000-0003-1035-3240}\inst{\ref{inst129}}\and
Hung-Yi Pu\orcid{0000-0001-9270-8812}\inst{\ref{inst136},\ref{inst137},\ref{inst11}}\and
Alexandra Rahlin\orcid{0000-0003-3953-1776}\inst{\ref{inst19}}\and
Venkatessh Ramakrishnan\orcid{0000-0002-9248-086X}\inst{\ref{inst16},\ref{inst138},\ref{inst139}}\and
Ramprasad Rao\orcid{0000-0002-1407-7944}\inst{\ref{inst10}}\and
Mark G. Rawlings\orcid{0000-0002-6529-202X}\inst{\ref{inst140},\ref{inst20},\ref{inst21}}\and
Angelo Ricarte\orcid{0000-0001-5287-0452}\inst{\ref{inst3},\ref{inst10}}\and
Luca Ricci\orcid{0000-0002-4175-3194}\inst{\ref{inst143}}\and
Bart Ripperda\orcid{0000-0002-7301-3908}\inst{\ref{inst131},\ref{inst144},\ref{inst132},\ref{inst26}}\and
Jan Röder\orcid{0000-0002-2426-927X}\inst{\ref{inst5}}\and
Freek Roelofs\orcid{0000-0001-5461-3687}\inst{\ref{inst29}}\and
Cristina Romero-Cañizales\orcid{0000-0001-6301-9073}\inst{\ref{inst11}}\and
Eduardo Ros\orcid{0000-0001-9503-4892}\inst{\ref{inst6}}\and
Arash Roshanineshat\orcid{0000-0002-8280-9238}\inst{\ref{inst14}}\and
Helge Rottmann\inst{\ref{inst6}}\and
Alan L. Roy\orcid{0000-0002-1931-0135}\inst{\ref{inst6}}\and
Ignacio Ruiz\orcid{0000-0002-0965-5463}\inst{\ref{inst123}}\and
Chet Ruszczyk\orcid{0000-0001-7278-9707}\inst{\ref{inst1}}\and
Kazi L. J. Rygl\orcid{0000-0003-4146-9043}\inst{\ref{inst109}}\and
León D. S. Salas\orcid{0000-0003-1979-6363}\inst{\ref{inst116}}\and
Salvador Sánchez\orcid{0000-0002-8042-5951}\inst{\ref{inst123}}\and
David Sánchez-Argüelles\orcid{0000-0002-7344-9920}\inst{\ref{inst75},\ref{inst76}}\and
Miguel Sánchez-Portal\orcid{0000-0003-0981-9664}\inst{\ref{inst123}}\and
Mahito Sasada\orcid{0000-0001-5946-9960}\inst{\ref{inst145},\ref{inst79},\ref{inst146}}\and
Kaushik Satapathy\orcid{0000-0003-0433-3585}\inst{\ref{inst14}}\and
Saurabh\orcid{0000-0001-7156-4848}\inst{\ref{inst6}}\and
Tuomas Savolainen\orcid{0000-0001-6214-1085}\inst{\ref{inst147},\ref{inst139},\ref{inst6}}\and
F. Peter Schloerb\inst{\ref{inst30}}\and
Jonathan Schonfeld\orcid{0000-0002-8909-2401}\inst{\ref{inst10}}\and
Karl-Friedrich Schuster\orcid{0000-0003-2890-9454}\inst{\ref{inst25}}\and
Lijing Shao\orcid{0000-0002-1334-8853}\inst{\ref{inst84},\ref{inst6}}\and
Zhiqiang Shen (\cntext{沈志强})\orcid{0000-0003-3540-8746}\inst{\ref{inst40},\ref{inst41}}\and
Sasikumar Silpa\orcid{0000-0003-0667-7074}\inst{\ref{inst16}}\and
Des Small\orcid{0000-0003-3723-5404}\inst{\ref{inst96}}\and
Randall Smith\orcid{0000-0003-4284-4167}\inst{\ref{inst10}}\and
Bong Won Sohn\orcid{0000-0002-4148-8378}\inst{\ref{inst42},\ref{inst94},\ref{inst43}}\and
Jason SooHoo\orcid{0000-0003-1938-0720}\inst{\ref{inst1}}\and
Kamal Souccar\orcid{0000-0001-7915-5272}\inst{\ref{inst30}}\and
Joshua S. Stanway\orcid{0009-0003-7659-4642}\inst{\ref{inst148}}\and
He Sun (\cntext{孙赫})\orcid{0000-0003-1526-6787}\inst{\ref{inst149},\ref{inst150}}\and
Fumie Tazaki\orcid{0000-0003-0236-0600}\inst{\ref{inst151}}\and
Alexandra J. Tetarenko\orcid{0000-0003-3906-4354}\inst{\ref{inst152}}\and
Remo P. J. Tilanus\orcid{0000-0002-6514-553X}\inst{\ref{inst14},\ref{inst29},\ref{inst90},\ref{inst153}}\and
Michael Titus\orcid{0000-0001-9001-3275}\inst{\ref{inst1}}\and
Kenji Toma\orcid{0000-0002-7114-6010}\inst{\ref{inst154},\ref{inst155}}\and
Pablo Torne\orcid{0000-0001-8700-6058}\inst{\ref{inst123},\ref{inst6}}\and
Teresa Toscano\orcid{0000-0003-3658-7862}\inst{\ref{inst5}}\and
Efthalia Traianou\orcid{0000-0002-1209-6500}\inst{\ref{inst5},\ref{inst6}}\and
Tyler Trent\inst{\ref{inst14}}\and
Sascha Trippe\orcid{0000-0003-0465-1559}\inst{\ref{inst156},\ref{inst157}}\and
Matthew Turk\orcid{0000-0002-5294-0198}\inst{\ref{inst45}}\and
Ilse van Bemmel\orcid{0000-0001-5473-2950}\inst{\ref{inst158}}\and
Huib Jan van Langevelde\orcid{0000-0002-0230-5946}\inst{\ref{inst96},\ref{inst90},\ref{inst159}}\and
Daniel R. van Rossum\orcid{0000-0001-7772-6131}\inst{\ref{inst29}}\and
Jesse Vos\orcid{0000-0003-3349-7394}\inst{\ref{inst160}}\and
Jan Wagner\orcid{0000-0003-1105-6109}\inst{\ref{inst6}}\and
Derek Ward-Thompson\orcid{0000-0003-1140-2761}\inst{\ref{inst148}}\and
John Wardle\orcid{0000-0002-8960-2942}\inst{\ref{inst161}}\and
Jasmin E. Washington\orcid{0000-0002-7046-0470}\inst{\ref{inst14}}\and
Jonathan Weintroub\orcid{0000-0002-4603-5204}\inst{\ref{inst10},\ref{inst3}}\and
Robert Wharton\orcid{0000-0002-7416-5209}\inst{\ref{inst6}}\and
Kaj Wiik\orcid{0000-0002-0862-3398}\inst{\ref{inst162},\ref{inst138},\ref{inst139}}\and
Gunther Witzel\orcid{0000-0003-2618-797X}\inst{\ref{inst6}}\and
Michael F. Wondrak\orcid{0000-0002-6894-1072}\inst{\ref{inst29},\ref{inst163}}\and
George N. Wong\orcid{0000-0001-6952-2147}\inst{\ref{inst164},\ref{inst35}}\and
Jompoj Wongphexhauxsorn\orcid{0000-0002-7730-4956}\inst{\ref{inst143},\ref{inst6}}\and
Qingwen Wu (\cntext{吴庆文})\orcid{0000-0003-4773-4987}\inst{\ref{inst165}}\and
Nitika Yadlapalli\orcid{0000-0003-3255-4617}\inst{\ref{inst22}}\and
Paul Yamaguchi\orcid{0000-0002-6017-8199}\inst{\ref{inst10}}\and
Aristomenis Yfantis\orcid{0000-0002-3244-7072}\inst{\ref{inst29}}\and
Doosoo Yoon\orcid{0000-0001-8694-8166}\inst{\ref{inst116}}\and
André Young\orcid{0000-0003-0000-2682}\inst{\ref{inst29}}\and
Ziri Younsi\orcid{0000-0001-9283-1191}\inst{\ref{inst166},\ref{inst47}}\and
Wei Yu (\cntext{于威})\orcid{0000-0002-5168-6052}\inst{\ref{inst10}}\and
Feng Yuan (\cntext{袁峰})\orcid{0000-0003-3564-6437}\inst{\ref{inst167}}\and
Ye-Fei Yuan (\cntext{袁业飞})\orcid{0000-0002-7330-4756}\inst{\ref{inst168}}\and
Ai-Ling Zeng (\cntext{曾艾玲})\orcid{0009-0000-9427-4608}\inst{\ref{inst5}}\and
J. Anton Zensus\orcid{0000-0001-7470-3321}\inst{\ref{inst6}}\and
Shuo Zhang\orcid{0000-0002-2967-790X}\inst{\ref{inst169}}\and
Guang-Yao Zhao\orcid{0000-0002-4417-1659}\inst{\ref{inst6},\ref{inst5}}\and
Shan-Shan Zhao (\cntext{赵杉杉})\orcid{0000-0002-9774-3606}\inst{\ref{inst40}}
}
\institute{
Steward Observatory and Department of Astronomy, University of Arizona, 933 N. Cherry Ave., Tucson, AZ 85721, USA\label{inst14}\and
Center for Astrophysics $|$ Harvard \& Smithsonian, 60 Garden Street, Cambridge, MA 02138, USA\label{inst10}\and
Black Hole Initiative at Harvard University, 20 Garden Street, Cambridge, MA 02138, USA\label{inst3}\and
Canadian Institute for Theoretical Astrophysics, University of Toronto, 60 St. George Street, Toronto, ON M5S 3H8, Canada\label{inst131}\and
Max-Planck-Institut für Radioastronomie, Auf dem Hügel 69, D-53121 Bonn, Germany\label{inst6}\and
Department of Astrophysics, Institute for Mathematics, Astrophysics and Particle Physics (IMAPP), Radboud University, P.O. Box 9010, 6500 GL Nijmegen, The Netherlands\label{inst29}\and
School of Space Research, Kyung Hee University, 1732, Deogyeong-daero, Giheung-gu, Yongin-si, Gyeonggi-do 17104, Republic of Korea\label{inst130}\and
Institute of Astronomy and Astrophysics, Academia Sinica, 11F of Astronomy-Mathematics Building, AS/NTU No. 1, Sec. 4, Roosevelt Rd., Taipei 106216, Taiwan, R.O.C.\label{inst11}\and
Institut für Theoretische Physik, Goethe-Universität Frankfurt, Max-von-Laue-Straße 1, D-60438 Frankfurt am Main, Germany\label{inst47}\and
Mizusawa VLBI Observatory, National Astronomical Observatory of Japan, 2-12 Hoshigaoka, Mizusawa, Oshu, Iwate 023-0861, Japan\label{inst79}\and
Graduate School of Science and Technology, Niigata University, 8050 Ikarashi 2-no-cho, Nishi-ku, Niigata 950-2181, Japan\label{inst101}\and
Astronomy Department, Universidad de Concepción, Casilla 160-C, Concepción, Chile\label{inst16}\and
Perimeter Institute for Theoretical Physics, 31 Caroline Street North, Waterloo, ON N2L 2Y5, Canada\label{inst26}\and
Department of Physics and Astronomy, University of Waterloo, 200 University Avenue West, Waterloo, ON N2L 3G1, Canada\label{inst27}\and
Waterloo Centre for Astrophysics, University of Waterloo, Waterloo, ON N2L 3G1, Canada\label{inst28}\and
Instituto de Astrofísica de Andalucía-CSIC, Glorieta de la Astronomía s/n, E-18008 Granada, Spain\label{inst5}\and
Massachusetts Institute of Technology Haystack Observatory, 99 Millstone Road, Westford, MA 01886, USA\label{inst1}\and
National Astronomical Observatory of Japan, 2-21-1 Osawa, Mitaka, Tokyo 181-8588, Japan\label{inst2}\and
Departament d'Astronomia i Astrofísica, Universitat de València, C. Dr. Moliner 50, E-46100 Burjassot, València, Spain\label{inst4}\and
Department of Physics, Faculty of Science, Universiti Malaya, 50603 Kuala Lumpur, Malaysia\label{inst7}\and
Department of Physics \& Astronomy, The University of Texas at San Antonio, One UTSA Circle, San Antonio, TX 78249, USA\label{inst8}\and
Physics \& Astronomy Department, Rice University, Houston, TX 77005-1827, USA\label{inst9}\and
Observatori Astronòmic, Universitat de València, C. Catedrático José Beltrán 2, E-46980 Paterna, València, Spain\label{inst12}\and
Department of Space, Earth and Environment, Chalmers University of Technology, Onsala Space Observatory, SE-43992 Onsala, Sweden\label{inst13}\and
Yale Center for Astronomy \& Astrophysics, Yale University, 52 Hillhouse Avenue, New Haven, CT 06511, USA\label{inst15}\and
Department of Physics, University of Illinois, 1110 West Green Street, Urbana, IL 61801, USA\label{inst17}\and
Fermi National Accelerator Laboratory, MS209, P.O. Box 500, Batavia, IL 60510, USA\label{inst18}\and
Department of Astronomy and Astrophysics, University of Chicago, 5640 South Ellis Avenue, Chicago, IL 60637, USA\label{inst19}\and
East Asian Observatory, 660 N. A'ohoku Place, Hilo, HI 96720, USA\label{inst20}\and
James Clerk Maxwell Telescope (JCMT), 660 N. A'ohoku Place, Hilo, HI 96720, USA\label{inst21}\and
California Institute of Technology, 1200 East California Boulevard, Pasadena, CA 91125, USA\label{inst22}\and
Institute of Astronomy and Astrophysics, Academia Sinica, 645 N. A'ohoku Place, Hilo, HI 96720, USA\label{inst23}\and
Department of Physics and Astronomy, University of Hawaii at Manoa, 2505 Correa Road, Honolulu, HI 96822, USA\label{inst24}\and
Institut de Radioastronomie Millimétrique (IRAM), 300 rue de la Piscine, F-38406 Saint Martin d'Hères, France\label{inst25}\and
Department of Astronomy, University of Massachusetts, Amherst, MA 01003, USA\label{inst30}\and
Instituto de Astronomia, Geofísica e Ciências Atmosféricas, Universidade de São Paulo, R. do Matão, 1226, São Paulo, SP 05508-090, Brazil\label{inst31}\and
Kavli Institute for Cosmological Physics, University of Chicago, 5640 South Ellis Avenue, Chicago, IL 60637, USA\label{inst32}\and
Department of Physics, University of Chicago, 5720 South Ellis Avenue, Chicago, IL 60637, USA\label{inst33}\and
Enrico Fermi Institute, University of Chicago, 5640 South Ellis Avenue, Chicago, IL 60637, USA\label{inst34}\and
Princeton Gravity Initiative, Jadwin Hall, Princeton University, Princeton, NJ 08544, USA\label{inst35}\and
Data Science Institute, University of Arizona, 1230 N. Cherry Ave., Tucson, AZ 85721, USA\label{inst36}\and
Program in Applied Mathematics, University of Arizona, 617 N. Santa Rita, Tucson, AZ 85721, USA\label{inst37}\and
Department of Physics, University of Maryland, 7901 Regents Drive, College Park, MD 20742, USA\label{inst38}\and
Cornell Center for Astrophysics and Planetary Science, Cornell University, Ithaca, NY 14853, USA\label{inst39}\and
Shanghai Astronomical Observatory, Chinese Academy of Sciences, 80 Nandan Road, Shanghai 200030, People's Republic of China\label{inst40}\and
Key Laboratory of Radio Astronomy and Technology, Chinese Academy of Sciences, A20 Datun Road, Chaoyang District, Beijing, 100101, People’s Republic of China\label{inst41}\and
Korea Astronomy and Space Science Institute, Daedeok-daero 776, Yuseong-gu, Daejeon 34055, Republic of Korea\label{inst42}\and
Department of Astronomy, Yonsei University, Yonsei-ro 50, Seodaemun-gu, 03722 Seoul, Republic of Korea\label{inst43}\and
WattTime, 490 43rd Street, Unit 221, Oakland, CA 94609, USA\label{inst44}\and
Department of Astronomy, University of Illinois at Urbana-Champaign, 1002 West Green Street, Urbana, IL 61801, USA\label{inst45}\and
Instituto de Astronomía, Universidad Nacional Autónoma de México (UNAM), Apdo Postal 70-264, Ciudad de México, México\label{inst46}\and
Institute of Astrophysics, Central China Normal University, Wuhan 430079, People's Republic of China\label{inst48}\and
Department of Astrophysical Sciences, Peyton Hall, Princeton University, Princeton, NJ 08544, USA\label{inst49}\and
NASA Hubble Fellowship Program, Einstein Fellow\label{inst50}\and
Dipartimento di Fisica ``E. Pancini'', Università di Napoli ``Federico II'', Compl. Univ. di Monte S. Angelo, Edificio G, Via Cinthia, I-80126, Napoli, Italy\label{inst51}\and
INFN Sez. di Napoli, Compl. Univ. di Monte S. Angelo, Edificio G, Via Cinthia, I-80126, Napoli, Italy\label{inst52}\and
Wits Centre for Astrophysics, University of the Witwatersrand, 1 Jan Smuts Avenue, Braamfontein, Johannesburg 2050, South Africa\label{inst53}\and
Department of Physics, University of Pretoria, Hatfield, Pretoria 0028, South Africa\label{inst54}\and
Centre for Radio Astronomy Techniques and Technologies, Department of Physics and Electronics, Rhodes University, Makhanda 6140, South Africa\label{inst55}\and
LESIA, Observatoire de Paris, Université PSL, CNRS, Sorbonne Université, Université de Paris, 5 place Jules Janssen, F-92195 Meudon, France\label{inst56}\and
JILA and Department of Astrophysical and Planetary Sciences, University of Colorado, Boulder, CO 80309, USA\label{inst57}\and
Tsung-Dao Lee Institute, Shanghai Jiao Tong University, Shengrong Road 520, Shanghai, 201210, People’s Republic of China\label{inst58}\and
National Astronomical Observatories, Chinese Academy of Sciences, 20A Datun Road, Chaoyang District, Beijing 100101, PR China\label{inst59}\and
Las Cumbres Observatory, 6740 Cortona Drive, Suite 102, Goleta, CA 93117-5575, USA\label{inst60}\and
Department of Physics, University of California, Santa Barbara, CA 93106-9530, USA\label{inst61}\and
National Radio Astronomy Observatory, 520 Edgemont Road, Charlottesville, VA 22903, USA\label{inst62}\and
Department of Electrical Engineering and Computer Science, Massachusetts Institute of Technology, 32-D476, 77 Massachusetts Ave., Cambridge, MA 02142, USA\label{inst63}\and
Google Research, 355 Main St., Cambridge, MA 02142, USA\label{inst64}\and
Institut für Theoretische Physik und Astrophysik, Universität Würzburg, Emil-Fischer-Str. 31, D-97074 Würzburg, Germany\label{inst65}\and
Department of History of Science, Harvard University, Cambridge, MA 02138, USA\label{inst66}\and
Department of Physics, Harvard University, Cambridge, MA 02138, USA\label{inst67}\and
NCSA, University of Illinois, 1205 W. Clark St., Urbana, IL 61801, USA\label{inst68}\and
Dipartimento di Fisica, Università degli Studi di Cagliari, SP Monserrato-Sestu km 0.7, I-09042 Monserrato (CA), Italy\label{inst69}\and
INAF - Osservatorio Astronomico di Cagliari, via della Scienza 5, I-09047 Selargius (CA), Italy\label{inst70}\and
INFN, sezione di Cagliari, I-09042 Monserrato (CA), Italy\label{inst71}\and
Institute for Mathematics and Interdisciplinary Center for Scientific Computing, Heidelberg University, Im Neuenheimer Feld 205, Heidelberg 69120, Germany\label{inst72}\and
Institut f\"ur Theoretische Physik, Universit\"at Heidelberg, Philosophenweg 16, 69120 Heidelberg, Germany\label{inst73}\and
CP3-Origins, University of Southern Denmark, Campusvej 55, DK-5230 Odense, Denmark\label{inst74}\and
Instituto Nacional de Astrofísica, Óptica y Electrónica. Apartado Postal 51 y 216, 72000. Puebla Pue., México\label{inst75}\and
Consejo Nacional de Humanidades, Ciencia y Tecnología, Av. Insurgentes Sur 1582, 03940, Ciudad de México, México\label{inst76}\and
Key Laboratory for Research in Galaxies and Cosmology, Chinese Academy of Sciences, Shanghai 200030, People's Republic of China\label{inst77}\and
Graduate School of Science, Nagoya City University, Yamanohata 1, Mizuho-cho, Mizuho-ku, Nagoya, 467-8501, Aichi, Japan\label{inst78}\and
Department of Physics, McGill University, 3600 rue University, Montréal, QC H3A 2T8, Canada\label{inst80}\and
Trottier Space Institute at McGill, 3550 rue University, Montréal,  QC H3A 2A7, Canada\label{inst81}\and
NOVA Sub-mm Instrumentation Group, Kapteyn Astronomical Institute, University of Groningen, Landleven 12, 9747 AD Groningen, The Netherlands\label{inst82}\and
Department of Astronomy, School of Physics, Peking University, Beijing 100871, People's Republic of China\label{inst83}\and
Kavli Institute for Astronomy and Astrophysics, Peking University, Beijing 100871, People's Republic of China\label{inst84}\and
Department of Astronomical Science, The Graduate University for Advanced Studies (SOKENDAI), 2-21-1 Osawa, Mitaka, Tokyo 181-8588, Japan\label{inst85}\and
Department of Astronomy, Graduate School of Science, The University of Tokyo, 7-3-1 Hongo, Bunkyo-ku, Tokyo 113-0033, Japan\label{inst86}\and
The Institute of Statistical Mathematics, 10-3 Midori-cho, Tachikawa, Tokyo, 190-8562, Japan\label{inst87}\and
Department of Statistical Science, The Graduate University for Advanced Studies (SOKENDAI), 10-3 Midori-cho, Tachikawa, Tokyo 190-8562, Japan\label{inst88}\and
Kavli Institute for the Physics and Mathematics of the Universe, The University of Tokyo, 5-1-5 Kashiwanoha, Kashiwa, 277-8583, Japan\label{inst89}\and
Leiden Observatory, Leiden University, Postbus 2300, 9513 RA Leiden, The Netherlands\label{inst90}\and
ASTRAVEO LLC, PO Box 1668, Gloucester, MA 01931, USA\label{inst91}\and
Applied Materials Inc., 35 Dory Road, Gloucester, MA 01930, USA\label{inst92}\and
Institute for Astrophysical Research, Boston University, 725 Commonwealth Ave., Boston, MA 02215, USA\label{inst93}\and
University of Science and Technology, Gajeong-ro 217, Yuseong-gu, Daejeon 34113, Republic of Korea\label{inst94}\and
National Institute of Technology, Ichinoseki College, Takanashi, Hagisho, Ichinoseki, Iwate, 021-8511, Japan\label{inst95}\and
Joint Institute for VLBI ERIC (JIVE), Oude Hoogeveensedijk 4, 7991 PD Dwingeloo, The Netherlands\label{inst96}\and
CSIRO, Space and Astronomy, PO Box 76, Epping, NSW 1710, Australia\label{inst97}\and
Department of Physics, Ulsan National Institute of Science and Technology (UNIST), Ulsan 44919, Republic of Korea\label{inst98}\and
Department of Physics, Korea Advanced Institute of Science and Technology (KAIST), 291 Daehak-ro, Yuseong-gu, Daejeon 34141, Republic of Korea\label{inst99}\and
Kogakuin University of Technology \& Engineering, Academic Support Center, 2665-1 Nakano, Hachioji, Tokyo 192-0015, Japan\label{inst100}\and
Physics Department, National Sun Yat-Sen University, No. 70, Lien-Hai Road, Kaosiung City 80424, Taiwan, R.O.C.\label{inst102}\and
Department of Astronomy, Kyungpook National University, 80 Daehak-ro, Buk-gu, Daegu 41566, Republic of Korea\label{inst103}\and
School of Astronomy and Space Science, Nanjing University, Nanjing 210023, People's Republic of China\label{inst104}\and
Key Laboratory of Modern Astronomy and Astrophysics, Nanjing University, Nanjing 210023, People's Republic of China\label{inst105}\and
INAF-Istituto di Radioastronomia, Via P. Gobetti 101, I-40129 Bologna, Italy\label{inst106}\and
Common Crawl Foundation, 9663 Santa Monica Blvd. 425, Beverly Hills, CA 90210 USA\label{inst107}\and
Instituto de Física, Pontificia Universidad Católica de Valparaíso, Casilla 4059, Valparaíso, Chile\label{inst108}\and
INAF-Istituto di Radioastronomia \& Italian ALMA Regional Centre, Via P. Gobetti 101, I-40129 Bologna, Italy\label{inst109}\and
Department of Physics, National Taiwan University, No. 1, Sec. 4, Roosevelt Rd., Taipei 106216, Taiwan, R.O.C\label{inst110}\and
Instituto de Radioastronomía y Astrofísica, Universidad Nacional Autónoma de México, Morelia 58089, México\label{inst111}\and
David Rockefeller Center for Latin American Studies, Harvard University, 1730 Cambridge Street, Cambridge, MA 02138, USA\label{inst112}\and
Yunnan Observatories, Chinese Academy of Sciences, 650011 Kunming, Yunnan Province, People's Republic of China\label{inst113}\and
Center for Astronomical Mega-Science, Chinese Academy of Sciences, 20A Datun Road, Chaoyang District, Beijing, 100012, People's Republic of China\label{inst114}\and
Key Laboratory for the Structure and Evolution of Celestial Objects, Chinese Academy of Sciences, 650011 Kunming, People's Republic of China\label{inst115}\and
Anton Pannekoek Institute for Astronomy, University of Amsterdam, Science Park 904, 1098 XH, Amsterdam, The Netherlands\label{inst116}\and
Gravitation and Astroparticle Physics Amsterdam (GRAPPA) Institute, University of Amsterdam, Science Park 904, 1098 XH Amsterdam, The Netherlands\label{inst117}\and
Center for Gravitation, Cosmology and Astrophysics, Department of Physics, University of Wisconsin–Milwaukee, P.O. Box 413, Milwaukee, WI 53201, USA\label{inst118}\and
Deceased\label{inst119}\and
School of Physics and Astronomy, Shanghai Jiao Tong University, 800 Dongchuan Road, Shanghai, 200240, People’s Republic of China\label{inst120}\and
SCOPIA Research Group, University of the Balearic Islands, Dept. of Mathematics and Computer Science, Ctra. Valldemossa, Km 7.5, Palma 07122, Spain\label{inst121}\and
Artificial Intelligence Research Institute of the Balearic Islands (IAIB), Palma 07122, Spain\label{inst122}\and
Institut de Radioastronomie Millimétrique (IRAM), Avenida Divina Pastora 7, Local 20, E-18012, Granada, Spain\label{inst123}\and
National Institute of Technology, Hachinohe College, 16-1 Uwanotai, Tamonoki, Hachinohe City, Aomori 039-1192, Japan\label{inst124}\and
Research Center for Astronomy, Academy of Athens, Soranou Efessiou 4, 115 27 Athens, Greece\label{inst125}\and
Department of Physics, Villanova University, 800 Lancaster Avenue, Villanova, PA 19085, USA\label{inst126}\and
Physics Department, Washington University, CB 1105, St. Louis, MO 63130, USA\label{inst127}\and
Departamento de Matemática da Universidade de Aveiro and Centre for Research and Development in Mathematics and Applications (CIDMA), Campus de Santiago, 3810-193 Aveiro, Portugal\label{inst128}\and
School of Physics, Georgia Institute of Technology, 837 State St NW, Atlanta, GA 30332, USA\label{inst129}\and
Dunlap Institute for Astronomy and Astrophysics, University of Toronto, 50 St. George Street, Toronto, ON M5S 3H4, Canada\label{inst132}\and
Canadian Institute for Advanced Research, 180 Dundas St West, Toronto, ON M5G 1Z8, Canada\label{inst133}\and
Dipartimento di Fisica, Università di Trieste, I-34127 Trieste, Italy\label{inst134}\and
INFN Sez. di Trieste, I-34127 Trieste, Italy\label{inst135}\and
Department of Physics, National Taiwan Normal University, No. 88, Sec. 4, Tingzhou Rd., Taipei 116, Taiwan, R.O.C.\label{inst136}\and
Center of Astronomy and Gravitation, National Taiwan Normal University, No. 88, Sec. 4, Tingzhou Road, Taipei 116, Taiwan, R.O.C.\label{inst137}\and
Finnish Centre for Astronomy with ESO, University of Turku, FI-20014 Turun Yliopisto, Finland\label{inst138}\and
Aalto University Metsähovi Radio Observatory, Metsähovintie 114, FI-02540 Kylmälä, Finland\label{inst139}\and
Gemini Observatory/NSF NOIRLab, 670 N. A’ohōkū Place, Hilo, HI 96720, USA\label{inst140}\and
Frankfurt Institute for Advanced Studies, Ruth-Moufang-Strasse 1, D-60438 Frankfurt, Germany\label{inst141}\and
School of Mathematics, Trinity College, Dublin 2, Ireland\label{inst142}\and
Julius-Maximilians-Universität Würzburg, Fakultät für Physik und Astronomie, Institut für Theoretische Physik und Astrophysik, Lehrstuhl für Astronomie, Emil-Fischer-Str. 31, D-97074 Würzburg, Germany\label{inst143}\and
Department of Physics, University of Toronto, 60 St. George Street, Toronto, ON M5S 1A7, Canada\label{inst144}\and
Department of Physics, Tokyo Institute of Technology, 2-12-1 Ookayama, Meguro-ku, Tokyo 152-8551, Japan\label{inst145}\and
Hiroshima Astrophysical Science Center, Hiroshima University, 1-3-1 Kagamiyama, Higashi-Hiroshima, Hiroshima 739-8526, Japan\label{inst146}\and
Aalto University Department of Electronics and Nanoengineering, PL 15500, FI-00076 Aalto, Finland\label{inst147}\and
Jeremiah Horrocks Institute, University of Central Lancashire, Preston PR1 2HE, UK\label{inst148}\and
National Biomedical Imaging Center, Peking University, Beijing 100871, People’s Republic of China\label{inst149}\and
College of Future Technology, Peking University, Beijing 100871, People’s Republic of China\label{inst150}\and
Tokyo Electron Technology Solutions Limited, 52 Matsunagane, Iwayado, Esashi, Oshu, Iwate 023-1101, Japan\label{inst151}\and
Department of Physics and Astronomy, University of Lethbridge, Lethbridge, Alberta T1K 3M4, Canada\label{inst152}\and
Netherlands Organisation for Scientific Research (NWO), Postbus 93138, 2509 AC Den Haag, The Netherlands\label{inst153}\and
Frontier Research Institute for Interdisciplinary Sciences, Tohoku University, Sendai 980-8578, Japan\label{inst154}\and
Astronomical Institute, Tohoku University, Sendai 980-8578, Japan\label{inst155}\and
Department of Physics and Astronomy, Seoul National University, Gwanak-gu, Seoul 08826, Republic of Korea\label{inst156}\and
SNU Astronomy Research Center, Seoul National University, Gwanak-gu, Seoul 08826, Republic of Korea\label{inst157}\and
ASTRON, Oude Hoogeveensedijk 4, 7991 PD Dwingeloo, The Netherlands\label{inst158}\and
University of New Mexico, Department of Physics and Astronomy, Albuquerque, NM 87131, USA\label{inst159}\and
Centre for Mathematical Plasma Astrophysics, Department of Mathematics, KU Leuven, Celestijnenlaan 200B, B-3001 Leuven, Belgium\label{inst160}\and
Physics Department, Brandeis University, 415 South Street, Waltham, MA 02453, USA\label{inst161}\and
Tuorla Observatory, Department of Physics and Astronomy, University of Turku, FI-20014 Turun Yliopisto, Finland\label{inst162}\and
Radboud Excellence Fellow of Radboud University, Nijmegen, The Netherlands\label{inst163}\and
School of Natural Sciences, Institute for Advanced Study, 1 Einstein Drive, Princeton, NJ 08540, USA\label{inst164}\and
School of Physics, Huazhong University of Science and Technology, Wuhan, Hubei, 430074, People's Republic of China\label{inst165}\and
Mullard Space Science Laboratory, University College London, Holmbury St. Mary, Dorking, Surrey, RH5 6NT, UK\label{inst166}\and
Center for Astronomy and Astrophysics and Department of Physics, Fudan University, Shanghai 200438, People's Republic of China\label{inst167}\and
Astronomy Department, University of Science and Technology of China, Hefei 230026, People's Republic of China\label{inst168}\and
Department of Physics and Astronomy, Michigan State University, 567 Wilson Rd, East Lansing, MI 48824, USA\label{inst169}
}

  \abstract
  {In Very-Long Baseline Interferometric arrays, nearly co-located stations probe the largest scales and typically cannot resolve the observed source. In the absence of large-scale structure, closure phases constructed with these stations are zero and, since they are independent of station-based errors, they can be used to probe data issues. Here, we show with an expansion about co-located stations, how these trivial closure phases become non-zero with brightness distribution on smaller scales than their short baseline would suggest. When applied to sources that are made up of a bright compact and large-scale diffuse component, the trivial closure phases directly measure the centroid relative to the compact source and higher-order image moments. We present a technique to measure these image moments with minimal model assumptions and validate it on synthetic Event Horizon Telescope (EHT) data. We then apply this technique to 2017 and 2018 EHT observations of \m87 and find a weak preference for extended emission in the direction of the large-scale jet. We also apply it to 2021 EHT data and measure the source centroid about 1 mas northwest of the compact ring, consistent with the jet observed at lower frequencies.
}


   \maketitle

\section{Introduction} \label{sec:intro}

Very-long baseline interferometry (VLBI) is an observational technique in which unconnected telescopes can be computationally synthesized into an effective instrument with an aperture size equivalent to the distance between the telescopes \citep{TMS}. Each combination of stations is sensitive to a different image scale, inversely proportional to the baseline separation. Many VLBI sources are blazars, which resemble a collimated jet extending outward from a compact core \citep{MOJAVE2009,BU2022}. As such, robust imaging requires coverage on many baseline separation scales.

The Event Horizon Telescope (EHT) is a VLBI array observing at millimeter/sub-millimeter wavelengths with stations separated by up to the diameter of the Earth. Its resolution reaches tens of microarcseconds, making it possible to resolve the ring-like structures and shadow of two supermassive black holes, \m87 \citep{M87p1,M87p2,M87p3,M87p4,M87p5,M87p6,M87p7,M87p8,M87p9,M87_2018p1} and \sgra \citep{SgrA.p1,SgrA.p2,SgrA.p3,SgrA.p4,SgrA.p5,SgrA.p6,SgrA.p7,SgrA.p8}. It can also resolve the inner jet region of several other active galactic nuclei \citep{3C279_2020,CenA2021,J1924_2022,NRAO530_2023,3C84_2024,NGC1052_2024,Roder2025}. 

VLBI, and the EHT in particular, is plagued with single-station instrument systematics which significantly corrupt the signal. When the number of baselines exceeds the number of stations, it is possible to solve for a majority of these instrument systematics \citep{TMS}. However, these are solved for simultaneously with assumptions about the brightness distribution, which leads to difficulty in tying specific source structures to features in the data. It is possible to construct closure quantities, e.g., closure phases, closure amplitudes, and closure traces, in which a combination of data products is independent of these station-based instrument corruptions and more directly probe the source brightness distribution \citep{Jennison1958,Twiss1960,TMS,Blackburn2020,Broderick2020}.

\m87 has a prominent large-scale limb-brightened jet which extends for many arcseconds and is seen across the electromagnetic spectrum \citep{Curtis1918,Biretta1999,Perlman2007,Walker2018,Kim2018,EHTMWL2021,M87_RadioAstron_2023,EHTMWL2024,JWSTM872025}. Recent observations with the Global mm-VLBI Array (GMVA) at 3 mm reveal the jet connecting to the ring in \m87 \citep{Lu2023,Kim2025}. At 1.3 mm, observations with the Atacama Large Millimeter/submillimeter Array (ALMA) reconstruct the jet oriented about $288^\circ$ East of North and extending to angular scales of tens of arcseconds \citep{Goddi2021}.

The EHT array before 2021 was composed primarily of long baselines ($\gtrsim 2\ \text{G}\lambda$) and intrasite baselines ($\lesssim 2\ \text{M}\lambda$), leaving a significant gap in scales. Its observations of \m87 at 1.3 mm reveal an excess of flux density between the intrasite and next-shortest baselines \citep{M87p4,M87p6,M87_2018p1,M87_2021}. It is natural to identify the missing flux density with the jet, but this has not been robustly proven. Traditionally, imaging with the EHT involves removing the intrasite baselines, as they do not have sufficient coverage to image large-scale emission, but still contain enough flux density to potentially introduce image artifacts in the compact brightness distribution.

The EHT array in 2021 included two new stations, which improved coverage and added two sets of intermediate baselines in the range of 0.1-1 G$\lambda$ \citep{M87_2021}. These new baselines measure visibilities that are difficult to fit with only a compact source. Explorations of emission on scales of hundreds of microarcseconds suggest that image components along the jet can explain the emission on intermediate baselines, but details of the extended emission requires strong model assumptions \citep{saurabh2025}. Furthermore, ``trivial'' closure phases involving an intrasite baseline show a bias away from zero, an effect possible if significant large-scale structure were present, although this offset cannot be explained by the same emission suggested for intermediate baselines \citep{M87_2021}.

In this paper, we present a new method to extract information about the location and structure of large-scale emission from closure phases, without strong model assumptions. We focus on EHT data as applicable to \m87, localize the centroid of the source, and show that the excess systematic signal in closure phases is fully explainable by source structure. In \autoref{sec:derivation}, we introduce closure phases, expand them for short baselines, and connect them to large-scale image moments. In \autoref{sec:validation}, we construct a synthetic EHT \m87 dataset and validate our method for centroid extraction. In \autoref{sec:application}, we apply this technique to EHT observations of \m87 and conclude in \autoref{sec:conclusion}. In addition. we explore the effects of polarization and leakage in \autoref{sec:polarization}. In \autoref{sec:higherorder}, we comment on possibilities of measuring higher-order image modes, and provide fitting details in \autoref{sec:fittingdetails}.

\section{Closure Quantities on Large Scales} \label{sec:derivation}
For an interferometric array with $N$ stations, it is possible to construct a set of closure quantities, each of which are invariant to a particular type of single-station corruptions. Here, we introduce closure phases and expand them in the limit that they probe large-scale structure. A similar derivation for polarimetric closure phases is given in \autoref{sec:polarization}.

\subsection{Definitions}

An interferometric baseline between stations $A$ and $B$ measures a visibility, related to the Fourier transform of the on-sky brightness distribution,
\begin{equation}
    \tilde{I}_{AB}=\iint I(\vec{x})\exp\left(-2\pi i \vec{u}_{AB}\cdot \vec{x}\right)d^2\vec{x}, \label{eq:FT}
\end{equation}
where $I$ is the Stokes I image intensity, $\vec{u}_{AB}$ is the projected baseline vector, and $\vec{x}$ is the on-sky angular coordinate vector relative to some origin. However, imperfections of the instrument lead to station-based complex gains,
\begin{equation}
    \tilde{I}_{AB,\text{observed}}=g_A \tilde{I}_{AB} g_B^*.
\end{equation}
When the number of baselines exceeds the number of stations, it becomes somewhat possible to solve for these complex gains simultaneous with the brightness distribution. It is instead possible to construct combinations of the visibilities which contain partial information of the source independent of these instrument artifacts. Relevant for this work, a closure phase on triangle $ABC$ is defined as
\begin{equation}
    \psi_{ABC}=\text{Arg}\left(\tilde{I}_{AB}\tilde{I}_{BC}\tilde{I}_{CA}\right),
\end{equation}
in which all the gain terms drop out. While closure phases do allow for a corruption-free probe of the source, their uncertainties become highly non-gaussian at a low signal-to-noise ratio. All datasets used in this work have been scan-averaged and have a sufficiently high signal-to-noise ratio\footnote{Sub-optimal phase calibration can introduce baseline-based biases when averaging (see, e.g., \citealt{Marti-Vidal2008}).}.

As many VLBI sources are compact, the closure phases for triangles involving two co-located (intrasite) stations should be zero, and these ``trivial'' closure phases can be used to probe non-closing errors in the data \citep{M87p3}. This is only possible when the co-located baseline probes scales much larger than the source size. 

\subsection{Short-Baseline Expansions}

Let $A'$ be a station that is almost co-located with $A$, so we may expand in short $\vec{u}_{AA'}$. The visibilities are
\begin{equation}
    \tilde{I}_{AA'}\approx\mathcal{F}\left(1-2\pi i \vec{u}_{AA'}\cdot \vec{\mathcal{C}}\right)
\end{equation}
where the total flux and centroid of emission are
\begin{align}
    \mathcal{F}&=\iint I(\vec{x})d^2\vec{x},\\
    \mathcal{C}_i&=\frac{1}{\mathcal{F}}\iint I(\vec{x})x_id^2\vec{x}.
\end{align}
Furthermore, $\vec{u}_{A'B}=\vec{u}_{AB}
-\vec{u}_{AA'}$, so
\begin{equation}
    \tilde{I}_{A'B}\approx\tilde{I}_{AB}-\vec{u}_{AA'}\cdot\vec{\nabla}_u\tilde{I}_{AB}.
\end{equation}

Expanding the trivial closure phases, 
\begin{equation}
    \psi_{AA'B}\approx-2\pi\vec{u}_{AA'}\cdot \vec{\mathcal{C}}-\vec{u}_{AA'}\cdot\vec{\nabla}_u \text{Arg}\left(\tilde{I}_{AB}\right).\label{eq:CPapprox}
\end{equation}
These closure phases probe the centroid of emission of the source relative to some zero point defined by the second term. Note that this expansion preserves the translational invariance of closure phases, as a phase gradient equally enters into both terms.

As closure phases are anti-symmetric, the next order terms go as $|\vec{u}|^3$ and contain higher image moments. The exploration of higher-order terms is given in \autoref{sec:higherorder}.

\subsection{Applicability to EHT Sources}
Many VLBI sources are composed of a compact core and extended diffuse emission, with significant substructure in their jet regions, like limb-brightening, filaments, and lobes (see, e.g, \citealt{Giovannini2018}; \citealt{Fuentes2023}). \m87 in particular has a compact emission region of $\sim 70\ \mu$as, and a jet that extends for many arcseconds, which is comparable to the beam size of individual EHT dishes \citep{M87p1}. The jet of \m87 is only visible on one side of the compact source, creating a first-order image moment, and is limb-brightened and radially falls off in intensity, thus creating third-order image moments \citep{Walker2018}.

The EHT baselines longer than $\sim 2~\text{G}\lambda$ roughly probe scales smaller than $100~\mu$as, and cannot probe the large-scale jet, resolving it out. In 2021, there were three sets of baselines in increasing separation that are potentially short enough to separate the large-scale emission\footnote{See \citet{M87_2021} for a full description of the stations in the 2021 EHT array.}:
\begin{itemize}
    \item JCMT-SMA, 0.1 M$\lambda$, 2 as
    \item ALMA-APEX, 2 M$\lambda$, 100 mas
    \item Kitt Peak-SMT, 0.1 G$\lambda$, 2 mas.
\end{itemize}
These themselves span many orders of magnitude and do not sufficiently cover the $uv$-plane.

With the EHT, we can construct many long skinny triangles by choosing three stations, $A$, $A'$, and $B$, such that $A$ and $A'$ are nearly co-located and $B$ is far from them. Let us model the observed brightness distribution as a point source that dominates the signal on baselines $AB$ and an extended source seen on baselines $AA'$. This allows us to remove all terms with derivatives of $I_{AB}$ from \autoref{eq:CPapprox}, as the visibilities should not change much over a length $|\vec{u}_{AA'}|$ for most of the $uv$-domain. This is an acceptable approximation where
\begin{equation}
    \vec{\nabla}_u \tilde{I}_{AB}\ll \vec{\nabla}_u \tilde{I}_{AA'},
\end{equation}
to within the thermal and systematic noise of the instrument. It is possible for this assumption to fail where phases jump rapidly near nulls in the visibility amplitudes, although it is unlikely for this to happen for every station $B$. This decomposition may be thought of as phase-referencing to the emission on the $AB$ baseline, and assuming that all long baselines can be phase-referenced simultaneously. 

Under this assumption, we find that
\begin{equation}
    \psi_{AA'B}\approx-2\pi \vec{u}_{AA'}\cdot \vec{\mathcal{C}}.\label{eq:firstmoment}
\end{equation}

\section{Application to Synthetic Data}\label{sec:validation}

\autoref{eq:firstmoment} provides a mathematical relationship between the trivial closure phases and physical properties of the source. Since it is a linear model, it can straightforwardly be fit to any combination of trivial closure phases, although differences in calibration over time and between stations can unduly affect certain parameter extractions. In this section, we fit synthetic data to validate the ability to reconstruct image moments and determine how much of the estimated systematic error is caused by the source structure.

\begin{figure}
    \centering
    \includegraphics[width=\columnwidth]{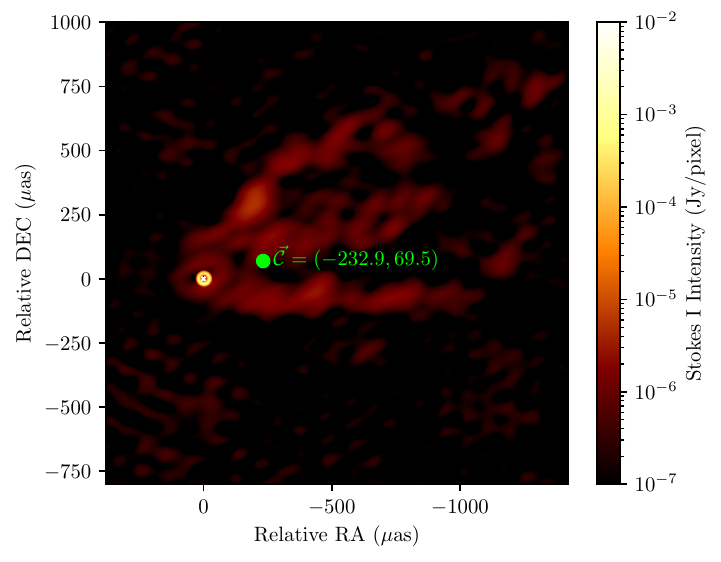}
    \caption{Synthetic dataset used for validation. The emission is composed of a bright compact ring and an extended jet. The green dot is the centroid of the image relative to the white x in the center of the ring.}
    \label{fig:GMVAcentroid}
\end{figure}

\autoref{fig:GMVAcentroid} shows an image designed to emulate \m87. It is composed of a bright ring structure and the extended jet from 2018 GMVA observations \citep{Lu2023}. Using this image as input and the software package \texttt{ehtim} \citep{ehtim_2018}, we create synthetic EHT data using the coverage and properties of the 2021 April 18 array at 227.1 GHz, the same dataset as used in \citet{M87_2021}. Following a necessity to model non-closing errors in EHT data, we inflate the error budget by $1\%$ of the visibility amplitudes, added in quadrature to the thermal errors. Many sources of non-closing errors are not explicitly included in the synthetic data generation pipelines, so this inflation serves mostly to match uncertainties on the real data and is expected to overestimate the synthetic data systematic errors.

\begin{figure}
    \centering
    \includegraphics[width=0.93\columnwidth]{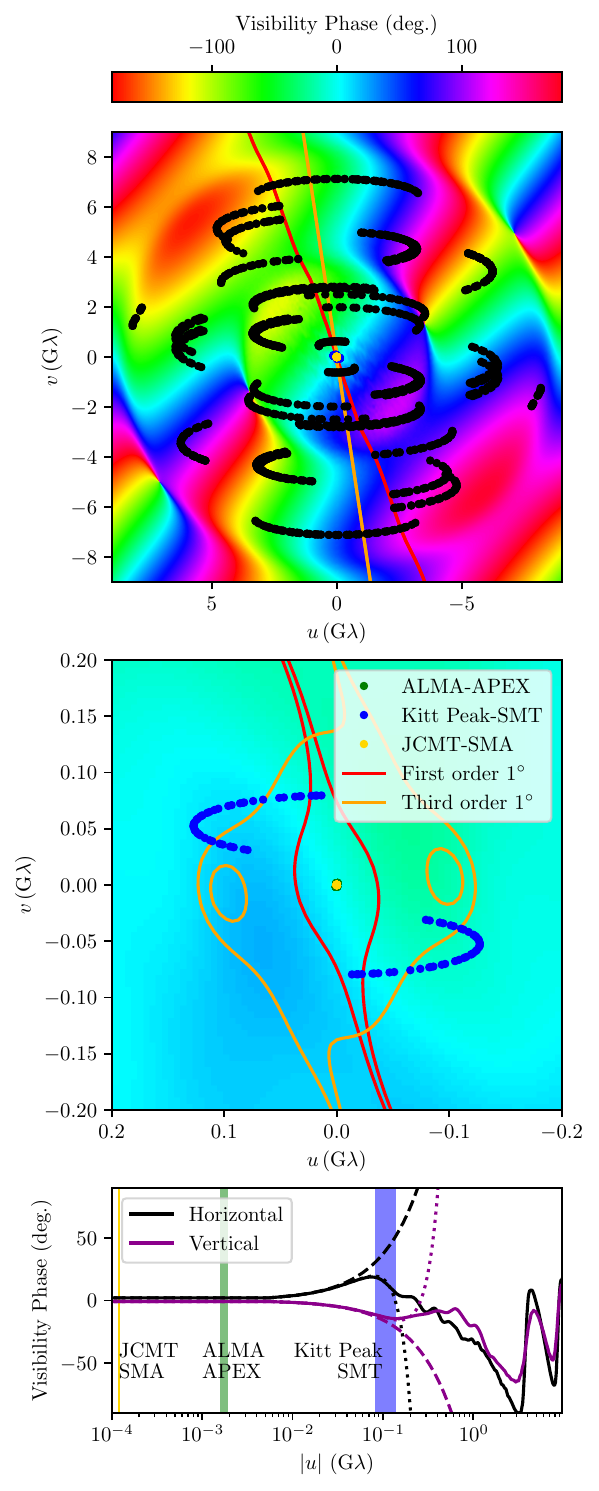}
    \caption{The top and middle panel show the visibility phases for the source model in \autoref{fig:GMVAcentroid}. The middle panel is a zoomed in version of the top panel. Black points show the $(u,v)$ locations of all synthetic observations, where the innermost three have been highlighted. The red and orange contours show the regions where, respectively, the first- and third-order approximations to the phases (\autoref{eq:firstmoment} and \autoref{eq:thirdmoments}, respectively) differ from the true phases by less than 1 degree. The bottom panel shows a horizontal and vertical slice of the phases as well as the first- and third- order approximations as dashed and dotted lines, respectively.}
    \label{fig:phases_contour}
\end{figure}

We first check whether the phase-centering and higher-order terms in the closure phase expansion are small compared to the centroid term. \autoref{fig:phases_contour} shows a plot of the visibility phases corresponding to the input brightness distribution. In order to tie large-scale closure quantities to the visibility phases, we require that the phase gradients with respect to $\vec{u}$ at long baselines are small compared to the phase gradients near zero. That is, we require that the trivial closure phases approximate the visibility phase of the short baseline in the trivial triangle. We find that under an appropriate choice of image center, phases wrap over several G$\lambda$, about a factor of 10 lower than phase gradients at short baselines. Note that to get small phase gradients at long baselines, the choice of phase center is located about 30 $\mu$as SE of the ring, and we cannot phase-center every baseline simultaneously. An important implication is that a centroid measured from the trivial closure phases is relative to a zero point not straightforwardly identifiable with the compact source, and introduces an additional uncertainty. However, we will find for this validation dataset that this additional uncertainty in the phase center relative to the compact ring will be smaller than the statistical uncertainty of the measurement of the centroid position offset.

The contours show the regions where the first- and third-order approximations to the phases agree to within 1 degree, similar to the uncertainty present in EHT observations. Within about 30 M$\lambda$, which encloses both the JCMT-SMA and ALMA-APEX baselines, the first-order approximation works, and we expect closure phases on triangles involving these baselines to measure the centroid of emission. The Kitt Peak-SMT baseline requires at least a third-order approximation, and even then, there is an additional $\sim 5$ degree error at the $uv$-locations furthest from zero. Although this dataset has the Kitt Peak-SMT on the border of probing both the large- and small-scale structure, a larger more diffuse jet could push the region of an acceptable approximation inside of the baseline lengths probed, and vice versa for a smaller jet.

\subsection{Synthetic First-Order Fits}\label{sec:firstorder}

\begin{figure}
    \centering
    \includegraphics[width=\columnwidth]{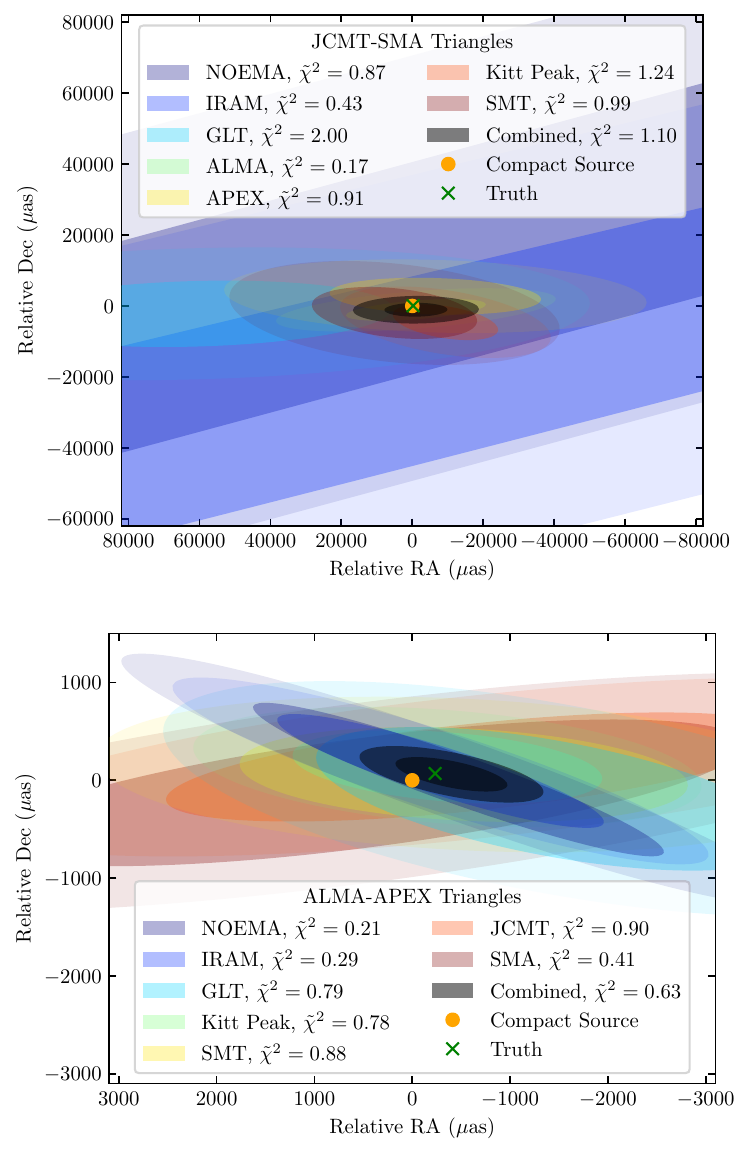}
    \caption{Covariance ellipses for the fits of the centroid position offset measured from synthetic data. The top panel shows triangles involving JCMT and SMA, while the bottom panel shows triangles involving ALMA and APEX. The black ellipses shows the 2-dimensional 68\% and 95\% confidence region from fits over the entire dataset, while all other colors split up the data into separate triangles. The orange dot is the phase reference, which is assumed to coincide with the compact ring, and the green x is the truth, with the same coordinates as in \autoref{fig:GMVAcentroid}. Stations are colored East to West (blue to red), which somewhat corresponds to the short baseline rotating over the course of a night. $\tilde{\chi}^2$ as defined in \autoref{sec:fittingdetails} is the reduced chi-squared statistic, and characterizes the goodness of fit. Note the bottom panel is zoomed relative to the top.}
    \label{fig:GMVAcentroidfit}
\end{figure}

We now fit the linear model to the trivial closure phases using the procedure described in \autoref{sec:fittingdetails}. \autoref{fig:GMVAcentroidfit} shows the extracted centroid position offset separately measured using JCMT-SMA and ALMA-APEX triangles. Fits are shown when including only one triangle and when all stations are included. A fit using only one data point at one time would create a linear band perpendicular to the direction of the intrasite baseline at that time. Since both JCMT-SMA and ALMA-APEX point roughly North-South, the uncertainty lies mainly in the East-West direction. As we include more times, the short baseline in the triangle rotates forming the covariance ellipses shown in the figure. In particular, ALMA-APEX-NOEMA and ALMA-APEX-IRAM are only seen in the beginning of the observations, while ALMA-APEX-JCMT and ALMA-APEX-SMA are only seen at the end. During this time, the ALMA-APEX baseline rotates by almost 45 degrees, thus breaking some of the degeneracy. 

Both sets of triangles agree with the true centroid location relative to the center of the ring (the green x lies within the black confidence regions in \autoref{fig:GMVAcentroidfit}). Every triangle is consistent with the combined fit and the truth value, indicating no significant unknown systematic errors. Due to the much shorter baseline and lower sensitivity of JCMT-SMA, these triangles cannot detect a centroid offset from zero, though they do constrain that the compound intensity distribution (i.e., the sum of the compact ring and extended component) is more compact than $\sim 10$ mas. The ALMA-APEX triangles do constrain both the direction and amplitude.

We had added 1\% fractional uncertainty to the synthetic data and have $\tilde{\chi}^2\lesssim 1$, suggesting that we have overinflated the extra uncertainty. The linear fits used here provide a technique to estimate the necessary fractional uncertainty by finding the amount required to get $\tilde{\chi}^2$ to unity, in essence linearly de-trending the contribution of large-scale structure to the error budget.    

\section{Application to \m87}\label{sec:application}

\begin{figure}
    \centering
    \includegraphics[width=\columnwidth]{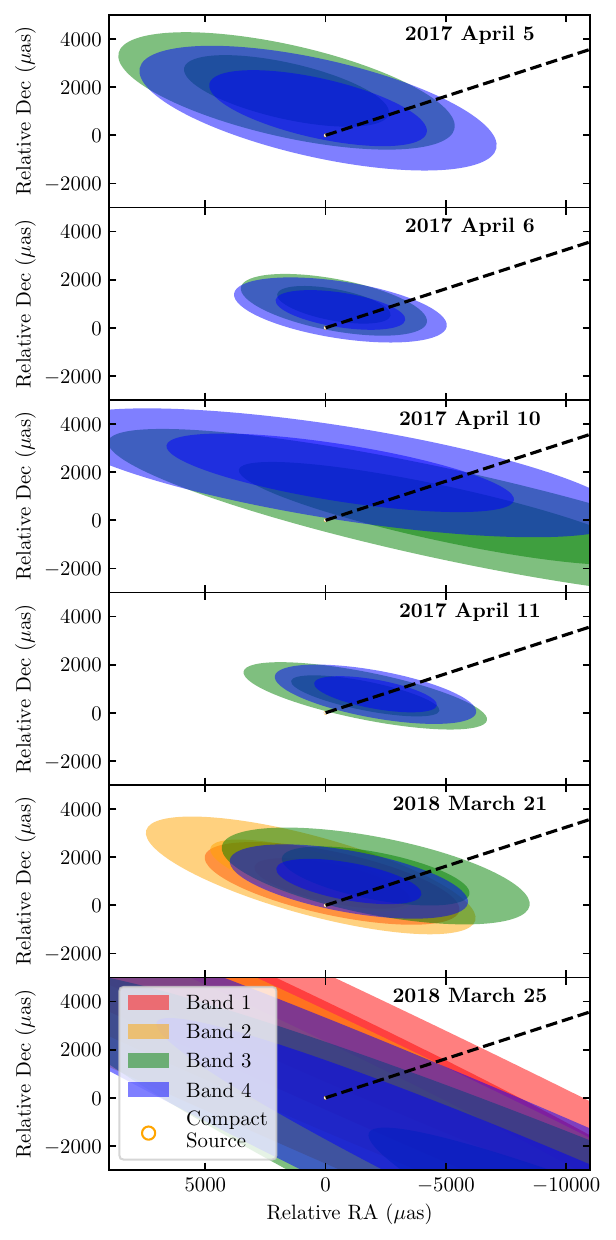}
    \caption{Estimated 2-dimensional 68\% and 95\% regions of the centroid position offset in \m87 for the 2017 and 2018 datasets. The black dashed line at $288^\circ$ East of North represents the direction of the mas-scale jet. }
    \label{fig:M87centroid1718}
\end{figure}

We now apply this method to \m87 data \citep{M87p1,M87_2018p1,M87_2021}. In 2017, the EHT observed \m87 on April 5, 6, 10, and 11 in two frequency bands centered on 227.1 and 229.1 GHz (band 3 and band 4, also called LO and HI). In 2018, observations with sufficient data were on March 21 and 25 with four frequency bands (bands 1-4, including those centered on 213.1 and 215.1 GHz). During these years, the only participating short baselines were ALMA-APEX and JCMT-SMA. In 2021, the EHT observed \m87 on April 13 and 18 in the four frequency bands, and added the short Kitt Peak-SMT baseline. To each dataset, we add $1\%$ fractional uncertainty to model systematic errors. Other data preparation steps match those in \citet{M87_2021}. Due to JCMT observing in 2017 and 2018 with only one polarization hand, we create closure phases for the whole array in the hand available for JCMT. In 2021, we create Stokes I closure phases.

We fit all JCMT-SMA and ALMA-APEX triangles simultaneously using the linear model with the same method as in \autoref{sec:validation}. The fit quality is listed in \autoref{table:chi2}.

\begin{table}
\caption{Goodness of fit measurements, $\tilde{\chi}^2$, for \m87 datasets.}             
\label{table:chi2}      
\centering    
\begin{tabular}{c c c c c}    
\hline\hline      
Dataset & Band 1 & Band 2 & Band 3 & Band 4 \\  
\hline                        
   2017 April 5 & - & - & 0.93 & 1.18 \\     
   2017 April 6 & - & - & 1.26 & 1.19 \\
   2017 April 10 & - & - & 0.84 & 0.56 \\
   2017 April 11 & - & - & 1.24 & 0.99 \\
   2018 March 21 & 1.24 & 0.75 & 1.39 & 0.89 \\ 
   2018 March 25 & 0.99 & 0.78 & 0.94 & 1.09 \\
   2021 April 13 & - & - & 1.1 & 1.02 \\
   2021 April 18 & - & - & 0.91 & 1.02 \\
\hline                              
\end{tabular}
\end{table}

\subsection{2017 and 2018}

\autoref{fig:M87centroid1718} shows the results of extracting the position offset of the centroid using the trivial triangles for 2017 and 2018 observations. There is no definitive constraint on a nonzero centroid offset, but the fits weakly suggest an excess of emission NW of the compact source located less than $\sim 4$ mas away. Furthermore, measurements are consistent across years and bands and each dataset is independently a good fit with a linear dependence on baseline length. This suggest that excesses in trivial closure phases are expected due to source structure, and not necessarily indicative of other systematic effects in the data. The recovered direction preference is consistent with the large-scale jet in \m87 which points roughly $288^\circ$ East of North.

\subsection{2021}

\begin{figure}
    \centering
    \includegraphics[width=\columnwidth]{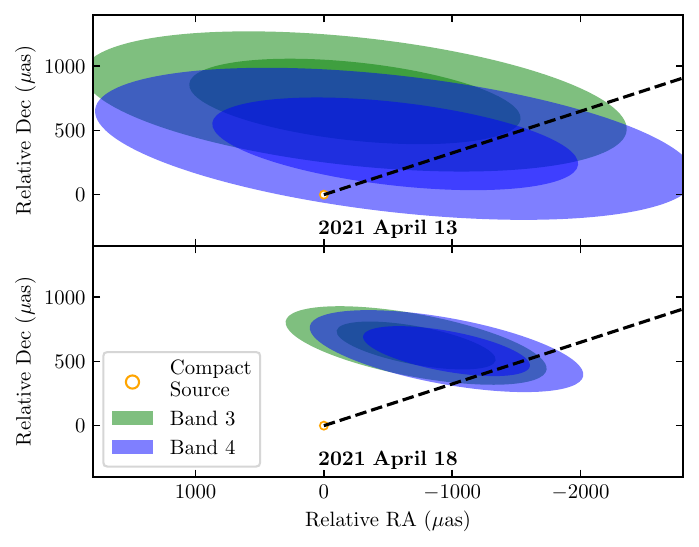}
    \caption{Estimated 2-dimensional 68\% and 95\% regions of the centroid position offset in \m87 from 2021 April 13 and 18 data, the latter of which significantly nonzero. The centroid is located about 1 mas Northwest of the compact source and consistent both between bands and with the direction of the large scale jet (dashed line).}
    \label{fig:M87centroid21}
\end{figure}

\autoref{fig:M87centroid21} shows the estimated centroid location using 2021 EHT data. Due to better sensitivity and more stations, the centroid can be more tightly constrained. In April 18 data, there is a clear offset located about 1 mas Northwest of the ring, consistent with observations of the jet at lower frequencies. The bands agree with each other, and all fits are statistically good, indicating no need for further systematic uncertainties or higher image moments.

\begin{figure}
    \centering
    \includegraphics[width=\columnwidth]{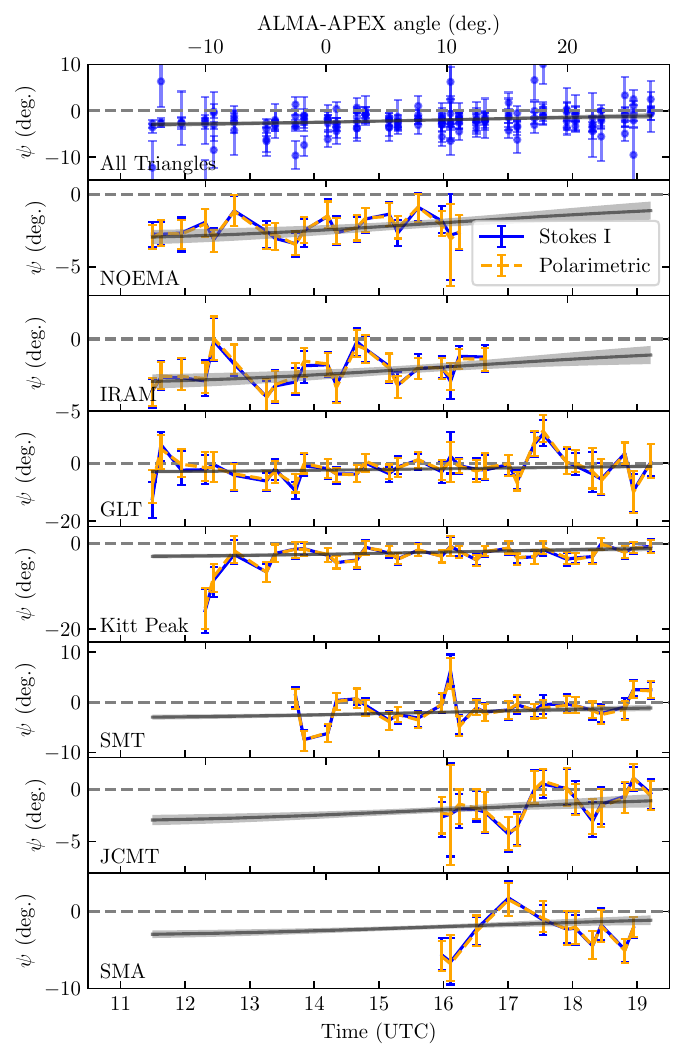}
    \caption{Closure phases on ALMA-APEX triangles from EHT 2021 April 18 data in band 3. The top panel shows all triangles and the other panels separate out each triangle. Stokes I closure phases are in blue and polarimetric closure phases are in orange. The black line and gray shaded region show the mean and 95\% fit region corresponding to \autoref{fig:M87centroid21}. The closure phases are offset from zero, consistently with one another, and well match the linear fit. The top ticks convert time to the ALMA-APEX baseline direction measured East of North.}
    \label{fig:fit_triangles_2021}
\end{figure}

\autoref{fig:fit_triangles_2021} shows the ALMA-APEX closure phases. They are systematically biased away from zero, but are all consistent with each other, hinting that the source of this bias is not station- or baseline-dependent. The linear fit matches this offset. The main driver for a nonzero closure phase comes from NOEMA and PV, since it is at this early time that the ALMA-APEX baseline is oriented close to the direction of the jet and thus being sensitive to the emission structure along the jet direction.

To explore whether leakage between polarization hands can be the source of the closure phase offset, we introduce polarimetric closure phases in \autoref{sec:polarization}, which are further invariant to any station-based corruption, including polarization leakage and differences between right and left gains. The Stokes I closure phases agree strongly with these polarimetric closure phases, indicating that station-based corruptions and source polarization are not biasing the centroid estimation. 

Closure phases on JCMT-SMA triangles and all other triangles, similarly show no serious signs of polarization leakage-based corruption. However, due to the lower sensitivity and smaller baseline length, the JCMT-SMA closure phases are consistent with zero and negligibly influence the measurements, and are included in the fits purely for consistency reasons. 

The centroid fits are not consistent with closure phases on Kitt Peak-SMT triangles. Following \autoref{sec:higherorder}, it may seem possible to include higher-order terms and more tightly constrain the centroid as well as higher-order image moments, which are then expected to inform on particulars of limb-brightening and intensity profiles. However, these closure phases are not fit well with third or higher orders, indicating that this baseline lies far outside of the regime where the expansion in \autoref{sec:derivation} holds. The parameter estimates, even of the centroid position offset, using this baseline could thus be significantly biased. A better interpretation for \m87 is that there is more structure on 100 $\mu$as-scales that is neither describable by a few image moments of the large-scale structure nor by a compact point source, such as that explored in \citet{saurabh2025}. 

It is similarly possible that the ALMA-APEX closure phases contain non-negligible higher-order image moments, as we are terminating their expansion at first order and creating a trivial degeneracy. This corresponds to unknown structure on baselines shorter than 2~M$\lambda$, like, for example, bright jet structure at large scales. This emission can be constrained by data from shorter ALMA-only baselines \citep{Goddi2021}, and whether its phases remain approximately linear with baseline length. Furthermore, particularly for interpreting higher-order moments, it becomes increasingly difficult to simultaneously reference all triangles (i.e., zero the increasing number of derivative terms in the expansion) as the small-scale ring structure further deviates from the point source assumption. Both of these effects are fundamental limitations of applying this method.

There is a further interpretational note that this centroid position offset measurement is of the combined jet and ring structure. More useful is the centroid only of the the diffuse jet, which is further away by a factor of the ratio of the jet flux divided by the total flux. For \m87, this value varies and is about $0.5\ \text{Jy}/1.5\ \text{Jy}$, so the jet centroid is expected to be 3 times further away. This is too close to the ring for the centroid position offset to be caused by the innermost bright jet component, HST-1 \citep{Biretta1999}, but it can be used to place limits on the total flux of it and other jet components at 230 GHz.

\section{Conclusions}\label{sec:conclusion}

We present a new technique to extract information about large-scale structure from interferometric closure phases applicable to sources with a compact core and large-scale diffuse emission. Triangles which involve co-located stations with a baseline that probes structure much larger than the spatial scales probed by longer baselines have a closure phase of zero. We expand these closure phases for baselines which are short, but potentially see offsets due to emission at large-scales. We find that, to first order, these trivial closure phases are directly proportional to the position offset of the centroid as measured relative to the compact source. The third-order components probe some combination of the first, second, and third moments of the total source brightness distribution. We thus create a linear model to extract these image moments from interferometric data, with few assumptions about the brightness distribution. This model is further invariant to a host of station-based signal corruptions.

We validate on a synthetic dataset composed of a bright compact ring and a large-scale diffuse jet designed to imitate EHT observations of \m87. We identify two potential sources of bias in the reconstructed moments. First, when phase gradients at long-baselines are large, the centroid (and higher moments) will be measured relative to a location not necessarily identifiable with that of the compact source. For sources similar to \m87, this is expected to add a subleading source of uncertainty. Second, the closure phases may be dominated by a higher-order term in baseline length than assumed, which can lead to formally good but biased fits. 

When only including triangles with intrasite baselines, we find that the sparse EHT coverage is sufficient to recover the centroid of the source. Each individual track contributes in the same direction, indicating that there is an informative non-zero signal present in the trivial closure phases. Longer intrasite baselines and those with a higher signal-to-noise ratio lead to tighter constraints. When including slightly longer baselines, the resulting localization of the centroid can tighten significantly and (combinations of) higher image moments can be recovered. However, this relies on an assumption that the short baseline visibilities are dominated by large-scale emission.

We apply this technique to EHT observations of \m87 in 2017, 2018, and 2021. In the first two years, there is weak evidence for non-zero trivial closure phases, which corresponds to extra emission Northwest of the ring. In 2021, the data strongly support a source centroid located about 1 milliarcsecond Northwest of the ring, consistent with jet direction measurements at lower frequencies. This detection is consistent among closure triangles, frequency bands, and is inconsistent with being caused by polarization leakage. Importantly, it solves the issue of nonzero closure phases on trivial triangles in 2021 EHT data. Poor quality of fits to longer baselines suggest that their data contain a significant component from compact structure. Better constraints on large-scale image moments would require more baseline coverage in the 10-100 M$\lambda$ range, such as those possible with the Korean VLBI Network. Some improvements can also be made with an intrasite baseline oriented East-West with similar sensitivity and separations as ALMA-APEX.

The method described here is applicable to a wide array of VLBI astronomy. With minimal source and instrument assumptions, it becomes possible to measure large-scale image moments, whose identification with specific source structure may require further assumptions. Specifically, this method can be applied to Centaurus A, 3C279, and 3C273, where ALMA observations at 230 GHz show directed extended emission \citep{Goddi2021}. It can also be used to identify what types of model components must be added to imaging algorithms to fit short baselines or, alternatively, which data would need to be removed in the imaging process. Furthermore, by detrending out structural effects, it becomes easier to identify what systematic biases in closure products are caused by correlation artifacts, and can be used as a more sophisticated model for network calibration \citep{M87p3, Blackburn2019}.

\begin{acknowledgements} 
The Event Horizon Telescope Collaboration thanks the following
organizations and programs: the Academia Sinica; the Academy
of Finland (projects 274477, 284495, 312496, 315721); the Agencia Nacional de Investigaci\'{o}n 
y Desarrollo (ANID), Chile via NCN$19\_058$ (TITANs), Fondecyt 1221421 and BASAL FB210003; the Alexander
von Humboldt Stiftung (including the Feodor Lynen Fellowship); an Alfred P. Sloan Research Fellowship;
Allegro, the European ALMA Regional Centre node in the Netherlands, the NL astronomy
research network NOVA and the astronomy institutes of the University of Amsterdam, Leiden University, and Radboud University;
the ALMA North America Development Fund; the Astrophysics and High Energy Physics programme by MCIN (with funding from European Union NextGenerationEU, PRTR-C17I1); the Black Hole Initiative, which is funded by grants from the John Templeton Foundation (60477, 61497, 62286) and the Gordon and Betty Moore Foundation (Grant GBMF-8273) - although the opinions expressed in this work are those of the author and do not necessarily reflect the views of these Foundations; 
the Brinson Foundation; the Canada Research Chairs (CRC) program; Chandra DD7-18089X and TM6-17006X; the China Scholarship
Council; the China Postdoctoral Science Foundation fellowships (2020M671266, 2022M712084); ANID through Fondecyt Postdoctorado (project 3250762); Conicyt through Fondecyt Postdoctorado (project 3220195); Consejo Nacional de Humanidades, Ciencia y Tecnología (CONAHCYT, Mexico, projects U0004-246083, U0004-259839, F0003-272050, M0037-279006, F0003-281692, 104497, 275201, 263356, CBF2023-2024-1102, 257435); the Colfuturo Scholarship; the Consejo Superior de Investigaciones 
Cient\'{i}ficas (grant 2019AEP112);
the Delaney Family via the Delaney Family John A.
Wheeler Chair at Perimeter Institute; Dirección General de Asuntos del Personal Académico-Universidad Nacional Autónoma de México (DGAPA-UNAM, projects IN112820 and IN108324); the Dutch Research Council (NWO) for the VICI award (grant 639.043.513), the grant OCENW.KLEIN.113, and the Dutch Black Hole Consortium (with project No. NWA 1292.19.202) of the research programme the National Science Agenda; the Dutch National Supercomputers, Cartesius and Snellius  (NWO grant 2021.013); 
the EACOA Fellowship awarded by the East Asia Core
Observatories Association, which consists of the Academia Sinica Institute of Astronomy and Astrophysics, the National Astronomical Observatory of Japan, Center for Astronomical Mega-Science,
Chinese Academy of Sciences, and the Korea Astronomy and Space Science Institute; 
the European Research Council (ERC) Synergy Grant ``BlackHoleCam: Imaging the Event Horizon of Black Holes'' (grant 610058) and Synergy Grant ``BlackHolistic:  Colour Movies of Black Holes:
Understanding Black Hole Astrophysics from the Event Horizon to Galactic Scales'' (grant 10107164); 
the European Union Horizon 2020
research and innovation programme under grant agreements
RadioNet (No. 730562), 
M2FINDERS (No. 101018682); the European Research Council for advanced grant ``JETSET: Launching, propagation and 
emission of relativistic jets from binary mergers and across mass scales'' (grant No. 884631); the European Horizon Europe staff exchange (SE) programme HORIZON-MSCA-2021-SE-01 grant NewFunFiCO (No. 10108625); the Horizon ERC Grants 2021 programme under grant agreement No. 101040021; the FAPESP (Funda\c{c}\~ao de Amparo \'a Pesquisa do Estado de S\~ao Paulo) under grant 2021/01183-8; the Fondes de Recherche Nature et Technologies (FRQNT); the Fondo CAS-ANID folio CAS220010; the Generalitat Valenciana (grants APOSTD/2018/177 and  ASFAE/2022/018) and
GenT Program (project CIDEGENT/2018/021); the Gordon and Betty Moore Foundation (GBMF-3561, GBMF-5278, GBMF-10423);   
the Institute for Advanced Study; the ICSC – Centro Nazionale di Ricerca in High Performance Computing, Big Data and Quantum Computing, funded by European Union – NextGenerationEU; the Istituto Nazionale di Fisica
Nucleare (INFN) sezione di Napoli, iniziative specifiche
TEONGRAV; 
the International Max Planck Research
School for Astronomy and Astrophysics at the
Universities of Bonn and Cologne; the Italian Ministry of University and Research (MUR)– Project CUP F53D23001260001, funded by the European Union – NextGenerationEU; 
Deutsche Forschungsgemeinschaft (DFG) research grant ``Jet physics on horizon scales and beyond'' (grant No. 443220636) and DFG research grant 443220636;
Joint Columbia/Flatiron Postdoctoral Fellowship (research at the Flatiron Institute is supported by the Simons Foundation); 
the Japan Ministry of Education, Culture, Sports, Science and Technology (MEXT; grant JPMXP1020200109); 
the Japan Society for the Promotion of Science (JSPS) Grant-in-Aid for JSPS
Research Fellowship (JP17J08829); the Joint Institute for Computational Fundamental Science, Japan; the Key Research
Program of Frontier Sciences, Chinese Academy of
Sciences (CAS, grants QYZDJ-SSW-SLH057, QYZDJSSW-SYS008, ZDBS-LY-SLH011); 
the Leverhulme Trust Early Career Research
Fellowship; the Max-Planck-Gesellschaft (MPG);
the Max Planck Partner Group of the MPG and the
CAS; the MEXT/JSPS KAKENHI (grants 18KK0090, JP21H01137,
JP18H03721, JP18K13594, 18K03709, JP19K14761, 18H01245, 25120007, 19H01943, 21H01137, 21H04488, 22H00157, 23K03453); the MICINN Research Projects PID2019-108995GB-C22, PID2022-140888NB-C22; the MIT International Science
and Technology Initiatives (MISTI) Funds; 
the Ministry of Science and Technology (MOST) of Taiwan (103-2119-M-001-010-MY2, 105-2112-M-001-025-MY3, 105-2119-M-001-042, 106-2112-M-001-011, 106-2119-M-001-013, 106-2119-M-001-027, 106-2923-M-001-005, 107-2119-M-001-017, 107-2119-M-001-020, 107-2119-M-001-041, 107-2119-M-110-005, 107-2923-M-001-009, 108-2112-M-001-048, 108-2112-M-001-051, 108-2923-M-001-002, 109-2112-M-001-025, 109-2124-M-001-005, 109-2923-M-001-001, 
110-2112-M-001-033, 110-2124-M-001-007 and 110-2923-M-001-001); the National Science and Technology Council (NSTC) of Taiwan
(111-2124-M-001-005, 112-2124-M-001-014,  112-2112-M-003-010-MY3, and 113-2124-M-001-008);
the Ministry of Education (MoE) of Taiwan Yushan Young Scholar Program;
the Physics Division, National Center for Theoretical Sciences of Taiwan;
the National Aeronautics and
Space Administration (NASA, Fermi Guest Investigator
grant 
80NSSC23K1508, NASA Astrophysics Theory Program grant 80NSSC20K0527, NASA NuSTAR award 
80NSSC20K0645); NASA Hubble Fellowship Program Einstein Fellowship;
NASA Hubble Fellowship 
grants HST-HF2-51431.001-A, HST-HF2-51482.001-A, HST-HF2-51539.001-A, HST-HF2-51552.001A awarded 
by the Space Telescope Science Institute, which is operated by the Association of Universities for 
Research in Astronomy, Inc., for NASA, under contract NAS5-26555; 
the National Institute of Natural Sciences (NINS) of Japan; the National
Key Research and Development Program of China
(grant 2016YFA0400704, 2017YFA0402703, 2016YFA0400702); the National Science and Technology Council (NSTC, grants NSTC 111-2112-M-001 -041, NSTC 111-2124-M-001-005, NSTC 112-2124-M-001-014); the US National
Science Foundation (NSF, grants AST-0096454,
AST-0352953, AST-0521233, AST-0705062, AST-0905844, AST-0922984, AST-1126433, OIA-1126433, AST-1140030,
DGE-1144085, AST-1207704, AST-1207730, AST-1207752, MRI-1228509, OPP-1248097, AST-1310896, AST-1440254, 
AST-1555365, AST-1614868, AST-1615796, AST-1715061, AST-1716327,  AST-1726637, 
OISE-1743747, AST-1743747, AST-1816420, AST-1935980, AST-1952099, AST-2034306,  AST-2205908, AST-2307887, AST-2407810); 
NSF Astronomy and Astrophysics Postdoctoral Fellowship (AST-1903847); 
the Natural Science Foundation of China (grants 11650110427, 10625314, 11721303, 11725312, 11873028, 11933007, 11991052, 11991053, 12192220, 12192223, 12273022, 12325302, 12303021); 
the Natural Sciences and Engineering Research Council of
Canada (NSERC); 
the National Research Foundation of Korea (the Global PhD Fellowship Grant: grants NRF-2015H1A2A1033752; the Korea Research Fellowship Program: NRF-2015H1D3A1066561; Brain Pool Program: RS-2024-00407499;  Basic Research Support Grant 2019R1F1A1059721, 2021R1A6A3A01086420, 2022R1C1C1005255, 2022R1F1A1075115); the POSCO Science Fellowship of the POSCO TJ Park Foundation; NOIRLab, which is managed by the Association of Universities for Research in Astronomy (AURA) under a cooperative agreement with the National Science Foundation; 
Onsala Space Observatory (OSO) national infrastructure, for the provisioning
of its facilities/observational support (OSO receives funding through the Swedish Research Council under grant 2017-00648);  the Perimeter Institute for Theoretical Physics (research at Perimeter Institute is supported by the Government of Canada through the Department of Innovation, Science and Economic Development and by the Province of Ontario through the Ministry of Research, Innovation and Science); the Portuguese Foundation for Science and Technology (FCT) grants (Individual CEEC program – 5th edition, CIDMA
through the FCT Multi-Annual Financing Program for R\&D Units UID/04106, CERN/FIS-PAR/0024/2021, 2022.04560.PTDC); the Princeton Gravity Initiative; the Spanish Ministerio de Ciencia, Innovaci\'{o}n  y Universidades (grants PID2022-140888NB-C21, PID2022-140888NB-C22, PID2023-147883NB-C21, RYC2023-042988-I); the Severo Ochoa grant CEX2021-001131-S funded by MICIU/AEI/10.13039/501100011033; The European Union’s Horizon Europe research and innovation program under grant agreement No. 101093934 (RADIOBLOCKS); The European Union “NextGenerationEU”, the Recovery, Transformation and Resilience Plan, the CUII of the Andalusian Regional Government and the Spanish CSIC through grant AST22\_00001\_Subproject\_10; ``la Caixa'' Foundation (ID 100010434) through fellowship codes LCF/BQ/DI22/11940027 and LCF/BQ/DI22/11940030; 
the University of Pretoria for financial aid in the provision of the new 
Cluster Server nodes and SuperMicro (USA) for a SEEDING GRANT approved toward these 
nodes in 2020; the Shanghai Municipality orientation program of basic research for international scientists (grant no. 22JC1410600); 
the Shanghai Pilot Program for Basic Research, Chinese Academy of Science, 
Shanghai Branch (JCYJ-SHFY-2021-013); the Simons Foundation (grant 00001470); the Spanish Ministry for Science and Innovation grant CEX2021-001131-S funded by MCIN/AEI/10.13039/501100011033; the Spinoza Prize SPI 78-409; the South African Research Chairs Initiative, through the 
South African Radio Astronomy Observatory (SARAO, grant ID 77948),  which is a facility of the National 
Research Foundation (NRF), an agency of the Department of Science and Innovation (DSI) of South Africa; the Swedish Research Council (VR); the Taplin Fellowship; the Toray Science Foundation; the UK Science and Technology Facilities Council (grant no. ST/X508329/1); the US Department of Energy (USDOE) through the Los Alamos National
Laboratory (operated by Triad National Security,
LLC, for the National Nuclear Security Administration
of the USDOE, contract 89233218CNA000001); and the YCAA Prize Postdoctoral Fellowship. This work was also supported by the National Research Foundation of Korea (NRF) grant funded by the Korea government(MSIT) (RS-2024-00449206). We acknowledge support from the Coordenação de Aperfeiçoamento de Pessoal de Nível Superior (CAPES) of Brazil through PROEX grant number 88887.845378/2023-00. We acknowledge financial support from Millenium Nucleus NCN23\_002 (TITANs) and Comité Mixto ESO-Chile.

We thank
the staff at the participating observatories, correlation
centers, and institutions for their enthusiastic support.
This paper makes use of the following ALMA data:
ADS/JAO.ALMA\#2017.1.00841.V and ADS/JAO.ALMA\#2019.1.01797.V.
ALMA is a partnership
of the European Southern Observatory (ESO;
Europe, representing its member states), NSF, and
National Institutes of Natural Sciences of Japan, together
with National Research Council (Canada), Ministry
of Science and Technology (MOST; Taiwan),
Academia Sinica Institute of Astronomy and Astrophysics
(ASIAA; Taiwan), and Korea Astronomy and
Space Science Institute (KASI; Republic of Korea), in
cooperation with the Republic of Chile. The Joint
ALMA Observatory is operated by ESO, Associated
Universities, Inc. (AUI)/NRAO, and the National Astronomical
Observatory of Japan (NAOJ). The NRAO
is a facility of the NSF operated under cooperative agreement
by AUI.
This research used resources of the Oak Ridge Leadership Computing Facility at the Oak Ridge National
Laboratory, which is supported by the Office of Science of the U.S. Department of Energy under contract
No. DE-AC05-00OR22725; the ASTROVIVES FEDER infrastructure, with project code IDIFEDER-2021-086; the computing cluster of Shanghai VLBI correlator supported by the Special Fund 
for Astronomy from the Ministry of Finance in China;  
We also thank the Center for Computational Astrophysics, National Astronomical Observatory of Japan. This work was supported by FAPESP (Fundacao de Amparo a Pesquisa do Estado de Sao Paulo) under grant 2021/01183-8.

APEX is a collaboration between the
Max-Planck-Institut f{\"u}r Radioastronomie (Germany),
ESO, and the Onsala Space Observatory (Sweden). The
SMA is a joint project between the SAO and ASIAA
and is funded by the Smithsonian Institution and the
Academia Sinica. The JCMT is operated by the East
Asian Observatory on behalf of the NAOJ, ASIAA, and
KASI, as well as the Ministry of Finance of China, Chinese
Academy of Sciences, and the National Key Research and Development
Program (No. 2017YFA0402700) of China
and Natural Science Foundation of China grant 11873028.
Additional funding support for the JCMT is provided by the Science
and Technologies Facility Council (UK) and participating
universities in the UK and Canada. 
The LMT is a project operated by the Instituto Nacional
de Astr\'{o}fisica, \'{O}ptica, y Electr\'{o}nica (Mexico) and the
University of Massachusetts at Amherst (USA).
The IRAM 30 m telescope on Pico Veleta, Spain and the NOEMA interferometer on Plateau de Bure,
France are operated by IRAM and supported by CNRS (Centre National de la Recherche Scientifique, France), MPG (Max-Planck-Gesellschaft, Germany), and IGN (Instituto Geográfico Nacional, Spain).
The SMT is operated by the Arizona
Radio Observatory, a part of the Steward Observatory
of the University of Arizona, with financial support of
operations from the State of Arizona and financial support
for instrumentation development from the NSF.
Support for SPT participation in the EHT is provided by the National Science Foundation through award OPP-1852617 
to the University of Chicago. Partial support is also 
provided by the Kavli Institute of Cosmological Physics at the University of Chicago. The SPT hydrogen maser was 
provided on loan from the GLT, courtesy of ASIAA.

This work used the
Extreme Science and Engineering Discovery Environment
(XSEDE), supported by NSF grant ACI-1548562,
and CyVerse, supported by NSF grants DBI-0735191,
DBI-1265383, and DBI-1743442. XSEDE Stampede2 resource
at TACC was allocated through TG-AST170024
and TG-AST080026N. XSEDE JetStream resource at
PTI and TACC was allocated through AST170028.
This research is part of the Frontera computing project at the Texas Advanced 
Computing Center through the Frontera Large-Scale Community Partnerships allocation
AST20023. Frontera is made possible by National Science Foundation award OAC-1818253.
This research was done using services provided by the OSG Consortium~\citep{osg07,osg09}, which is supported by the National Science Foundation award Nos. 2030508 and 1836650.
Additional work used ABACUS2.0, which is part of the eScience center at Southern Denmark University, and the Kultrun Astronomy Hybrid Cluster (projects Conicyt Programa de Astronomia Fondo Quimal QUIMAL170001, Conicyt PIA ACT172033, Fondecyt Iniciacion 11170268, Quimal 220002). 
Simulations were also performed on the SuperMUC cluster at the LRZ in Garching, 
on the LOEWE cluster in CSC in Frankfurt, on the HazelHen cluster at the HLRS in Stuttgart, 
and on the Pi2.0 and Siyuan Mark-I at Shanghai Jiao Tong University.
The computer resources of the Finnish IT Center for Science (CSC) and the Finnish Computing 
Competence Infrastructure (FCCI) project are acknowledged. This
research was enabled in part by support provided
by Compute Ontario (http://computeontario.ca), Calcul
Quebec (http://www.calculquebec.ca), and the Digital Research Alliance of Canada (https://alliancecan.ca/en).

The EHTC has
received generous donations of FPGA chips from Xilinx
Inc., under the Xilinx University Program. The EHTC
has benefited from technology shared under open-source
license by the Collaboration for Astronomy Signal Processing
and Electronics Research (CASPER). The EHT
project is grateful to T4Science and Microsemi for their
assistance with hydrogen masers. This research has
made use of NASA's Astrophysics Data System. We
gratefully acknowledge the support provided by the extended
staff of the ALMA, from the inception of
the ALMA Phasing Project through the observational
campaigns of 2017 and 2018. We would like to thank
A. Deller and W. Brisken for EHT-specific support with
the use of DiFX. We thank Martin Shepherd for the addition of extra features in the Difmap software 
that were used for the CLEAN imaging results presented in this paper.
We acknowledge the significance that
Maunakea, where the SMA and JCMT EHT stations
are located, has for the indigenous Hawaiian people.

\end{acknowledgements}

\bibliographystyle{aa}
\bibliography{references}{}
\appendix

\section{Polarization}\label{sec:polarization}

When the instrument contains effects related to instrumental polarization (also known as polarization leakage), the Stokes~I closure phases no longer probe the source brightness distribution directly. Here, we show that the method of expanding closure quantities at large scales can be similarly performed for a polarization-weighed centroid.

Each station in an interferometric array produces two measurements of the electric field, one for each polarization basis. For circularly polarized feeds, the correlations (i.e, visibilities) between the different polarization channels for each baseline can be arranged within a coherency matrix,
\begin{equation}
    \mathbf{V}_{AB}=
    \begin{pmatrix}
        RR_{AB} & RL_{AB}\\
        LR_{AB} & LL_{AB}
    \end{pmatrix},
\end{equation}
where $R$ and $L$ represent the right- and left-handed polarizations, respectively. Using the radio interferometer measurement equation \citep{Hamaker1996,Smirnov2011}, the elements of the coherency matrix can be identified with combinations of the Fourier transforms of the brightness distribution of the Stokes $I$, $Q$, $U$, and $V$ quantities,
\begin{equation}
    \mathbf{V}_{AB}=
    \begin{pmatrix}
        \tilde{I}_{AB}+\tilde{V}_{AB} & \tilde{Q}_{AB}+i\tilde{U}_{AB}\\
        \tilde{Q}_{AB}-i\tilde{U}_{AB} & \tilde{I}_{AB}-\tilde{V}_{AB}
    \end{pmatrix},
\end{equation}
where the Fourier Stokes quantities have a similar definition as in 
\autoref{eq:FT}. Station-based corruptions enter as
\begin{equation}
    \mathbf{V}_{AB,\text{observed}}=\mathbf{J}_A \mathbf{V}_{AB} \mathbf{J}_B^*,
\end{equation}
where $\mathbf{J}$ contain information about generic antenna-based instrumental and atmospheric effects (e.g., gains and polarization leakage). The corresponding polarimetric closure phase is\footnote{When comparing with Stokes I closure phases, there is a sign degeneracy when taking the square root. For short-baseline comparison in this work, it suffices to divide $\hat{\psi}$ by 2, but other triangles may need a shift of $180^\circ$.}
\begin{align}
    \hat{\psi}_{ABC}&=\frac{1}{2}\text{Arg}\left(\left|\left|\mathbf{V}_{AB}\mathbf{V}_{CB}^{-1}\mathbf{V}_{CA}\mathbf{V}_{BA}^{-1}\mathbf{V}_{BC}\mathbf{V}_{AC}^{-1}\right|\right|\right)\\
    &=\text{Arg}\left(||\mathbf{V}_{AB}\mathbf{V}_{BC}\mathbf{V}_{CA}||\right),
\end{align}
which is just a closure phase of the quantity $\tilde{I}^2-\tilde{Q}^2-\tilde{U}^2-\tilde{V}^2$. Performing the expansion in nearly co-located stations, we have
\begin{equation}
    \hat{\psi}_{AA'B}\approx -4 \pi \vec{u}_{AA'}\cdot \frac{\sum_s \vec{\mathcal{C}}_s \hat{\mathcal{F}}^2_s}{\sum_s \hat{\mathcal{F}}^2_s},
\end{equation}
where $s$ sums over the four Stokes quantities, $\vec{\mathcal{C}}_s$ are the centroids of the Stokes quantities, and $\hat{\mathcal{F}}^2=\{\mathcal{F}_I^2,-\mathcal{F}_Q^2,-\mathcal{F}_U^2,-\mathcal{F}_V^2\}$ are related to the square of their total fluxes. Evidently, the large-scale closure quantities are proportional to a polarization de-weighed centroid. For \m87, the polarization fraction is a few percent, and thus the centroid would be negligibly changed. Similarly, the Stokes $V$ flux is a few percent of the Stokes $I$ flux, thus the difference between using $RR$, $LL$, or Stokes $I$ closure phases is also negligible, unless significant polarization leakage is present.

The other three-station polarized closure quantity can be expressed as the magnitude of the trace,
\begin{equation}
    |\mathcal{T}_{ABC}|=\frac{1}{2}\left|\text{Tr}\left(\mathbf{V}_{AB}\mathbf{V}_{CB}^{-1}\mathbf{V}_{CA}\mathbf{V}_{BA}^{-1}\mathbf{V}_{BC}\mathbf{V}_{AC}^{-1}\right)\right|,
\end{equation}
and is equal to 1 when the source is unpolarized. For nearly co-located stations, 
\begin{equation}
    |\mathcal{T}_{AA'B}|\approx 1+4\pi^2\frac{\sum_{s,t} \left[\vec{u}_{AA'}\cdot(\vec{\mathcal{C}}_s-\vec{\mathcal{C}}_t)\right]^2 \hat{\mathcal{F}}^2_s\hat{\mathcal{F}}^2_t}{\left(\sum_s \hat{\mathcal{F}}^2_s\right)^2},
\end{equation}
and contains information about the distance between the centroids of the Stokes quantities. A constant polarization fraction and angle results in this trace having a magnitude of 1. To be non-unity, there must be a gradient of polarization relative to the Stokes $I$ emission. 

\section{Higher-Order Expansions}
\label{sec:higherorder}

\subsection{Third-Order Fits}

\label{sec:thirdorder}
The closure phases are anti-symmetric in $\vec{u}$ and thus contain only odd powers in their expansion. Expanding the derivation of the trivial closure phases in \autoref{sec:derivation} to third-order yields
\begin{align}
    \psi_{AA'B}\approx-2\pi \vec{u}_{AA'}\cdot \vec{\mathcal{C}}+\frac{(2\pi)^3}{3!}\left[\frac{\iint (\vec{u}_{AA'}\cdot\vec{x})^3 I(\vec{x})d^2\vec{x}}{\mathcal{F}}\right.&\nonumber\\
    \left.+2\left(\vec{u}_{AA'}\cdot \vec{\mathcal{C}}\right)^3-3\left(\vec{u}_{AA'}\cdot \vec{\mathcal{C}}\right)\frac{\iint (\vec{u}_{AA'}\cdot\vec{x})^2 I(\vec{x})d^2\vec{x}}{\mathcal{F}}\right]&\label{eq:thirdmoments}\\
    =-2\pi \vec{u}_{AA'}\cdot \vec{\mathcal{C}}+2\pi \psi^{(3)}_{ijk}u_{AA',i}u_{AA',j}u_{AA',k}&\nonumber.
\end{align}
Thus, the closure phases for triangles with short baselines probe some combination of both even and odd image moments.

\begin{figure*}
    \centering
    \includegraphics[width=\textwidth]{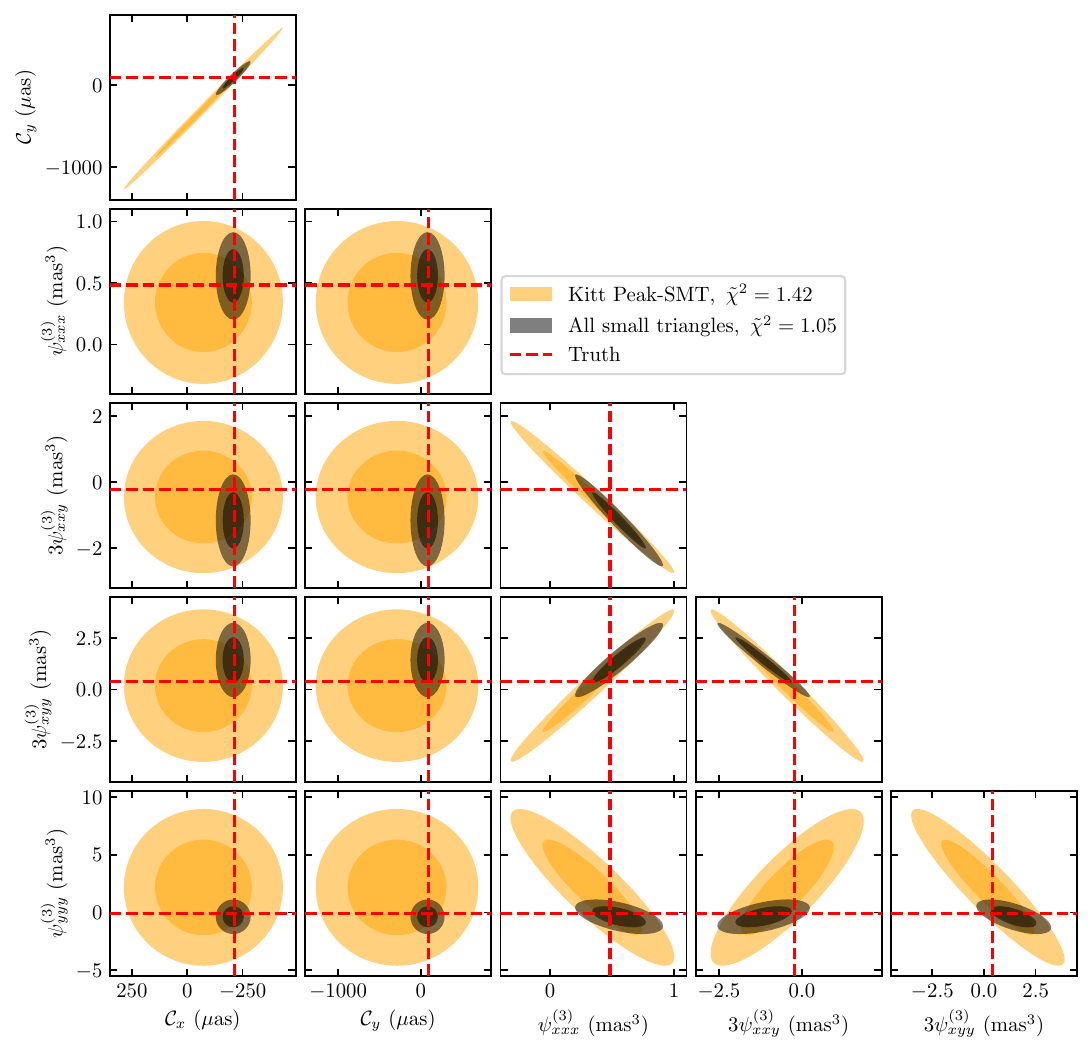}
    \caption{Two-dimensional 68\% and 95\% confidence regions of the third-order fits to the synthetic data set closure phases. Orange regions correspond to including only triangles with Kitt Peak-SMT and black regions include ALMA-APEX and JCMT-SMA triangles as well. Dashed red lines show the truth values measured with \autoref{eq:thirdmoments}. The truths are not recovered for all parameter combinations.}
    \label{fig:GMVAthirdorder}
\end{figure*}

\autoref{fig:GMVAthirdorder} shows the fits including a third-order component in baseline length. Since the Kitt Peak-SMT baselines have a significant East-West component, they can break the degeneracy in the mainly North-South first-order fits and narrow the centroid localization. Despite not having the high signal-to-noise ratio of the ALMA-APEX triangles, the much narrower (orange) bands stem from the longer Kitt Peak-SMT baseline length. When the additional information from the triangles involving ALMA-APEX and JCMT-SMA are added, we can tightly constrain the centroid location and modestly constrain some combination of the second- and third-order moments of the image. With further source assumptions, these could be tied to a measurement of the limb-brightening or a radial emission profile of the jet.

However, the fits do not necessarily cover the truth, particularly for the joint $\psi_{xxx}$-$\psi_{xxy}$ components. This is primarily due to the emergence of a fifth-order component we identified with \autoref{fig:phases_contour}. Although in the synthetic dataset used here, closure phases on Kitt Peak-SMT triangles happened be dominated by the large-scale signal, \m87 could have both more large-scale and more small-scale structures. Thus, interpretation of higher-order moments is more model-dependent than measuring a centroid, and will require a-priori knowledge of whether the Kitt Peak-SMT baselines are dominated by the diffuse jet or the compact source. Getting a statistically good fit to Kitt Peak-SMT triangles in \m87 2021 data requires much higher orders to the point that image moments are entirely unconstrained.

\subsection{Second order expansion}\label{sec:secondorder}

The expansions of the closure phases constrain some combination of the image moments, but it is not possible to disentangle each image moment with the information stored only in the closure phases. Although the closure amplitudes can be expanded the same way, they do not as straightforwardly map to large-scale image moments. This primarily happens because it is not possible to amplitude-reference to multiple long baselines similar to how we simultaneously phase-referenced to all long baselines before.

Instead, we can expand the visibility amplitudes directly and write
\begin{align}
    &|\tilde{I}_{AA',\text{observed}}|= |g_A||g_{A'}||\tilde{I}_{AA'}|\nonumber
    \\
    &\approx \mathcal{F}|g_A||g_{A'}|\left[1+2\pi^2 \left(\vec{u}_{AA'}\cdot\vec{\mathcal{C}}\right)^2-\iint \left(\vec{u}_{AA'}\cdot \vec{x}\right)^2I d^2\vec{x}\right]\label{eq:secondmoment}
\end{align}
Thus the visibility amplitudes contain the missing information necessary to convert the closure phase fits directly to image moments. However, we have now introduced an unmodelled term in the form of the gain amplitudes. 

\begin{figure}
    \centering
    \includegraphics[width=\columnwidth]{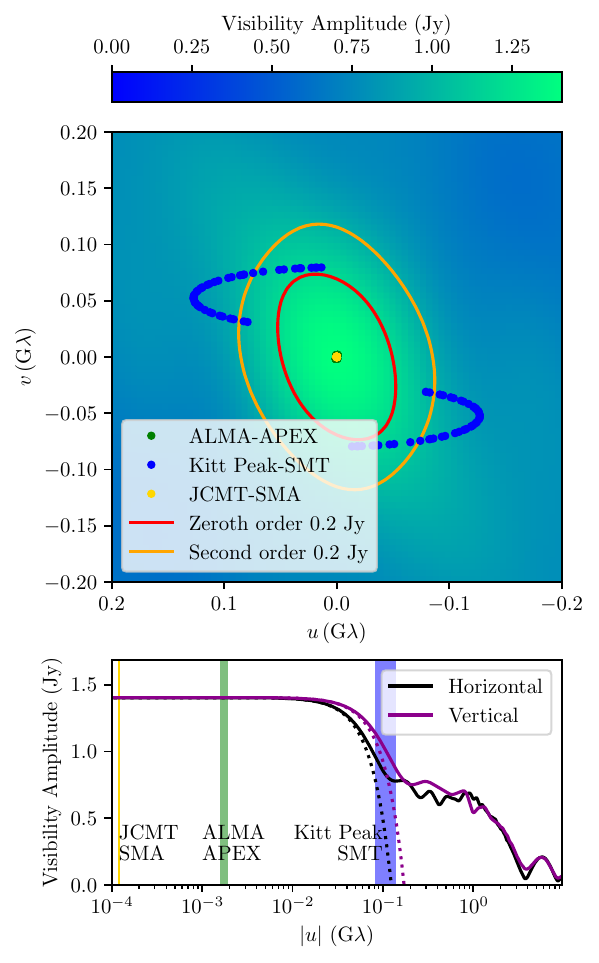}
    \caption{Visibility amplitudes for the source model in \autoref{fig:GMVAcentroid}. Green, blue, and gold points show the $(u,v)$ locations of for the shortest baselines. The red and orange contours show the regions where, respectively, the zeroth- and second-order approximations to the amplitudes differ from the true amplitudes by less than 0.2 Jy. The bottom panel shows horizontal and vertical slices of the amplitudes (solid lines) and the dotted lines show the second-order expansion.}
    \label{fig:amp_contour}
\end{figure}

\autoref{fig:amp_contour} shows a plot similar to \autoref{fig:phases_contour}, but for the visibility amplitudes. The contours show where the second order expansion from \autoref{eq:secondmoment} matches the input amplitudes to within 0.2 Jy. This cutoff is meant to mimic the maximum possible effect of the gain amplitudes. The expansion only works for the innermost Kitt Peak-SMT baselines, and for the points furthest from zero, the expansion almost reaches 0 Jy. An expansion in the logarithm of the amplitudes fares no better.

Fundamentally, this is because the amplitudes flatten with respect to $\vec{u}$ around 0.1 G$\lambda$ as they begin to have a significant component from the compact source. Higher-order expansions or an extra error term as in the third-order case are possible, but rapidly give diminishing returns. However, it seems that the location where the amplitudes reach the value of the compact flux and flatten is a good indicator of where the third-order closure phase expansion is expected to hold. 

\section{Fitting Details}
\label{sec:fittingdetails}

\autoref{eq:firstmoment}, \autoref{eq:thirdmoments}, and \autoref{eq:secondmoment} are all linear models of the form
\begin{equation}
    \vec{\psi}=A(\vec{u})\cdot\vec{x}
\end{equation}
where $\vec{x}$ depends on the centroid position offset and higher image moments, while the matrix $A$ contains the information about the baselines. As such, we can fit the model analytically. Let $\Omega$ be a diagonal matrix of the closure phase variances. The best fit values are given by
\begin{equation}
    \vec{x}=\left(A^T\Omega^{-1}A\right)^{-1}A^T\Omega^{-1}\vec{\psi}
\end{equation}
with a covariance of
\begin{equation}
    \Sigma=\left(A^T\Omega^{-1}A\right)^{-1}
\end{equation}
and a reduced chi-squared statistic of
\begin{equation}
    \tilde{\chi}^2=\frac{1}{N}\left(\vec{\psi}-A\vec{x}\right)\Omega^{-1}\left(\vec{\psi}-A\vec{x}\right),
\end{equation}
where $N$ is the number of measurements minus the number of parameters.
Where applicable, we show the 2-dimensional confidence region,
\begin{equation}
    n_\text{eff}=\sqrt{-2\ln \left[1-\text{erf}\left(\frac{n}{\sqrt{2}}\right)\right]}.
\end{equation}.

\end{document}